\documentclass[twocolumn,aps,prc]{revtex4}

\usepackage[english]{babel} 	
\usepackage{graphics} 
\usepackage[pdftex]{graphicx}
\usepackage{amssymb} 			
\usepackage{amsmath} 			
\usepackage{booktabs}
\usepackage{tabularx}
\usepackage{rotating}
\usepackage{enumerate}

\usepackage{float}
\usepackage{url}	
\usepackage{bm}

\graphicspath{{./figures/}}

\allowdisplaybreaks

\begin{document}

\title{ Confinement of two-body systems and calculations in $d$ dimensions.}

\author{E. Garrido$^{1}$}
\author{A.S. Jensen$^{2}$}

\affiliation{$^{1}$Instituto de Estructura de la Materia, IEM-CSIC,
Serrano 123, E-28006 Madrid, Spain  \\
$^{2}$Department of Physics and Astronomy, Aarhus University, DK-8000 Aarhus C, Denmark} 

\date{\today}

\begin{abstract}
A continuous transition for a system moving in a three-dimensional (3D) space to moving in a lower-dimensional 
space, 2D or 1D, can be made by means of an external squeezing potential. A squeeze along one direction 
gives rise to a 3D to 2D transition, whereas a simultaneous squeeze along two directions produces
a 3D to 1D transition, without going through an intermediate 2D configuration. In the same way, for a
system moving in a 2D space, a squeezing potential along one direction produces a 
2D to 1D transition. In this work we investigate the equivalence between this kind of confinement procedure
 and calculations without an external field, but where the dimension $d$ is taken as a parameter that changes 
 continuously from $d=3$ to $d=1$. The practical case of an external harmonic oscillator squeezing potential acting 
 on  a two-body system is investigated in details. For the three transitions considered, 3D~$\rightarrow$~2D, 2D~$\rightarrow$~1D,
 and 3D~$\rightarrow$~1D, a universal connection between the harmonic oscillator parameter and the dimension $d$ is found.
 This relation is well established for infinitely large 3D scattering lengths of the two-body potential for 3D~$\rightarrow$~2D
 and 3D~$\rightarrow$~1D transitions, and for infinitely large 2D scattering length for the 2D~$\rightarrow$~1D case.
 For finite scattering lengths size corrections must be applied.  The traditional wave functions for external squeezing potentials are
 shown to be uniquely related with the wave functions for specific
 non-integer dimension parameters, $d$.
\end{abstract}


\maketitle

\section{Introduction}

The properties of quantum systems depend crucially on the dimension of the space where they are allowed
to move.  A clear example of this is the centrifugal barrier in the radial Schr\"{o}dinger 
equation, which, for zero total angular momentum, is negative
for a two-body system in two (2D) dimensions, while it is zero in
three (3D) dimensions \cite{nie01}.  An immediate consequence of this is  that any infinitesimal amount of 
attraction produces a bound state in 2D, whereas in 3D a finite amount of attraction is necessary for 
binding a system \cite{sim79}. As an exotic example of recent interest we can mention, at the three-body level, the
occurrence in 3D of the Efimov effect \cite{efi70}. In 2D, this effect does not occur, neither for equal mass three-body 
systems \cite{bru79} nor for unequal mass systems \cite{lim80}. 

The transition from a three-dimensional to a lower-dimensional space is commonly
investigated by means of an external trap potential that confines the system
under investigation in a certain region in the space. In this way, in \cite{pet01}
a harmonic oscillator trap potential is used, and binary atomic collisions are
investigated under different confinement regimes. This work focus on scattering properties
in confined spaces, not necessarily similar to the asymmetric squeezing of only one or two spatial
dimensions.  More recently, the particle physics formalism has been specifically extended to connect 
$d=3$ and $d=2$ by continuously compactifying, by means of an infinite potential well, one of the
dimensions \cite{san18,ros18,yam15}. Working in momentum space, the intentions were to study three-body
physics, but two-body subsystems are then necessary ingredients. The physical interpretation of the 
squeezing parameter in this procedure is a problem necessary to be addressed to connect properly to 
measurements. The same compactifying procedure along one or two directions has been employed to investigate 
the $S$-matrix for two-body scattering \cite{bea19}, and 
again the focus is on scattering under confining conditions.
Another structure-related investigation has appeared in the literature, that is the superfluid phase transition 
temperature in the crossover from three to two dimensions \cite{lam16}.
This is necessarily a many-body effect although prompted by two-body properties.

In all these works the procedure has been to perform genuine 3D calculations where
the external potential enters explicitly in order to limit the space available. However, an alternative can be 
to employ an abstract formulation where the dimension $d$ can take 
different values describing the different possible scenarios. For instance, in \cite{dor86} an expansion in 
terms of $1/d$ is performed, where $d$ is thought of 
as an integer, allowing extrapolations between integers. The philosophy has been to extrapolate obtainable 
results as function of $1/d$ for very large $d$ and first down to $d=3$ for two or more particles \cite{gon91}.  
Going further down towards $d=2$ is probably going too far  \cite{gon91}, both because $1/d=1/2$ is not very large, 
but especially because the properties change qualitatively from $d=3$ to $d=2$.
Non-integer dimensions have
also been employed in various subfields of mathematics and physics, see
e.g. \cite{pes95,val12}.  For many
particles even mixed dimensions have been used to study exotic
structures [9]. A practical
continuous connection between integer dimensions is interesting in order
to understand
the related structure variations.


In this work the approach would be formulation in coordinate space by
simple analytic continuation of the abstract formulation in terms of
the dimension parameter $d$ assuming non-integer values.  The limits
of $d=1,2,3$ are now well defined in contrast to any value between
these integers.  A practical interpretation can be found by use of
deformed external fields squeezing one or more dimensions to zero
spatial extension corresponding to infinitely high zero point energy.
This method was used in a recent work \cite{gar19}, where the
continuous confinement of quantum systems from three to two dimensions
was investigated.

The confinement in \cite{gar19} was treated by use of two different
procedures. In the first one the particles are put under the effect of
an external trap potential acting on a single direction. This
potential continuously limits the motion of the particles along that
direction, in such a way that for infinite squeezing the system moves
in a 2D space. In the second method the external potential is not
used, and the problem instead is solved directly in $d$ dimensions,
where $d$ is a parameter that changes continuously from 3 to 2. This
formulation has the advantage that the numerical effort required is
similar to solving the ordinary problem for integer dimensions.

In \cite{gar19} the 3D to 2D confinement was investigated  for two-body systems and external harmonic oscillator 
confining potentials. The purpose of this work is to extend the investigation to squeezing up to one dimension (1D).
This can be done in two different ways.  In the first one we consider a simultaneous squeezing along two directions, in such
a way that the external potential pushes the system, initially moving in 3D, to moving in 1D  without going through 
an intermediate 2D geometry. The second procedure consists on two consecutive squeezing processes along one direction,
giving rise to a 3D to 2D squeezing followed by a 2D to 1D squeezing. The obvious, but complicated, extension to systems made of more than
two particles is left for a forthcoming work. 

In all the confinement scenarios (3D~$\rightarrow$~2D, 2D~$\rightarrow$~1D, and 3D~$\rightarrow$~1D) the problem will be 
treated by means of the two procedures described above, i.e., by explicit use of the confinement potential, and by use of 
the dimension $d$ as a parameter that changes continuously from $d=3$ to $d=2$ or $d=1$. One of the main goals is then,
for all the cases,  to establish the equivalence between a given value of the confining harmonic oscillator frequency and 
the dimension $d$ describing the same physical situation. 

The connection between the harmonic oscillator parameter and the
dimension $d$ should preferentially be universal in the sense of being
independent of the details of the potential. It is well known that necessary ingredients for the appearance of universal properties of quantum 
systems are the existence of two-body interactions with large scattering lengths,  and, to a large extent,
the preponderance of relative $s$-waves between the constituents. The existence, under these conditions, of a universal connection
between the harmonic oscillator parameter and the dimension will be investigated.  Here it is clear that large squeezing confining a wave function to be
inside the two-body potential must depend on potential details. However,  comparing the two methods,
it can still result in the same universal dependence, since both are subject to the same potential.  To be practical, we have 
established such a highly desirable connection between
the wave functions obtained in the two methods.

The overall purpose of the present work is therefore to study the a number of
different transitions between integer dimensions, and to establish the
universal connection between the $d$-parameter results and those of
the brute force three dimensional calculation with a deformed external
field.  The connection must allow the numerically simpler $d$-method
to be self-sufficient, that is in itself providing full information
including correspondence to external field and three dimensional wave
function.  The paper is organized as follows. In section~\ref{sec2} we
describe the procedure used to confine a two-body system by use of an
external harmonic oscillator potential.  In section~\ref{sec3} we
briefly describe the method used to solve the two-body problem in $d$
dimensions.  Section~\ref{app2} presents analytic results in the large
squeezing limit, that is close to one or two dimensions.
Sections~\ref{sec5} and ~\ref{sec6} present and discuss the numerical
results and section ~\ref{sec7} gives the universal translation
between the two methods.  Finally, Section~\ref{sec8} contains a summary
and the future perspectives are briefly discussed.
A mathematical connection between wave functions from the two methods are 
given in an appendix.

\section{Harmonic oscillator squeezing}
\label{sec2}

A simple way to confine particles is to put them under the effect of
an external potential with steep walls that forces them to move in a confined
space. Therefore, the problem to be solved is the usual
Schr\"{o}dinger equation, but where, together with the interaction
between the particles, the confining one-body potential has to be
included.

In this work we shall consider an external harmonic oscillator
potential whose frequency will be written as
\begin{equation}
\omega=\frac{\hbar}{m_\omega b_{ho}^2},
\label{eq1}
\end{equation}
where $m_\omega$ is some arbitrary mass.  Obviously, the smaller the
harmonic oscillator length $b_{ho}$, the more confined the particles
are in the corresponding direction.

In the following we describe how this harmonic oscillator potential is
treated for the three confinement cases, 3D~$\rightarrow$~2D,
2D~$\rightarrow$~1D, and 3D~$\rightarrow$~1D, considered in this work.

\subsection{3D~$\rightarrow$~2D}
\label{3a}

In this case the external harmonic oscillator potential is assumed to
act along the $z$-direction. Therefore, the problem to be solved here
will be the usual three-dimensional two-body problem but where, on top
of the two-body interaction, each of the two particles feels the
effect of the external trap potential:
\begin{equation}
V^{(i)}_{trap}=\frac{1}{2}m_i\omega^2 r_i^2 \cos^2\theta_i=\frac{1}{2} \frac{m_i \hbar^2}{m_\omega^2 b_{ho}^4} r_i^2 \cos^2\theta_i,
\label{eq2}
\end{equation}
where Eq.(\ref{eq1}) has been used, and where $r_i$ and $\theta_i$ are
the radial coordinate and polar angle associated to particle $i$ with
mass $m_i$.  Eventually, for $b_{ho}=0$ the particles can move only in
the two dimensions of the $xy$-plane.

As usual, the two-body wave function can be expanded in partial waves as:
\begin{equation}
\Psi(\bm{r})= \sum_{\ell m} \frac{u_\ell(r)}{r} Y_{\ell m}(\theta,\varphi),
\label{eq3}
\end{equation}
where $\bm{r} = \bm{r}_1 - \bm{r}_2$ is the relative coordinate between the two particles whose direction is given by the polar 
and azimuthal angles $\theta$ and $\varphi$, respectively. For simplicity, in the notation we shall assume 
spinless particles, although the generalization  to particles with non-zero spin is straightforward.

For two particles with masses $m_1$ and $m_2$ and coordinates $\bm{r}_1$ and $\bm{r}_2$ we have that:
\begin{equation}
r^2=\frac{m_1}{\mu}r_1^2+\frac{m_2}{\mu}r_2^2-\frac{m_1+m_2}{\mu}r_{cm}^2,
\end{equation}
where $\mu$ is the reduced mass and $\bm{r}_{cm}$ is the position of the two-body center of mass. This expression
permits us to write the full trap potential as:
\begin{eqnarray}
\lefteqn{
\frac{1}{2}m_1\omega^2r_1^2 \cos^2\theta_1+ \frac{1}{2}m_2\omega^2r_2^2\cos^2\theta_2=
}\nonumber \\  &&
\frac{1}{2}\mu\omega^2r^2\cos^2\theta +\frac{1}{2}(m_1+m_2) \omega^2 r_{cm}^2\cos^2\theta_{cm},
\end{eqnarray}
where $\theta_{cm}$ is the polar angle associated to $\bm{r}_{cm}$.
The expression above implies that, after removal of the center of mass motion, the squeezing potential to be used in the relative 
two-body calculation takes the form:
\begin{equation}
V_{trap}(r,\theta)=\frac{1}{2}\mu\omega^2r^2\cos^2\theta,
\label{trap}
\end{equation}
whose ground state energy is $E_{ho}=\hbar\omega/2$.

The wave functions $u_\ell$ in Eq.(\ref{eq3}) are the solutions of the radial Schr\"{o}dinger equation
\begin{eqnarray}
\lefteqn{ \hspace*{-1cm}
\left[ \frac{\partial^2}{\partial r^2}-\frac{\ell(\ell+1)}{r^2}-
\frac{2\mu}{\hbar^2} V_{2b}(r)+\frac{2\mu E_{tot}}{\hbar^2}
\right] u_\ell} \nonumber \\ && \hspace*{5mm}
-\frac{2\mu}{\hbar^2}\sum_{\ell' m'}\langle Y_{\ell m}|V_{trap}(r,\theta)|Y_{\ell' m'}\rangle_\Omega
                            u_{\ell'}=0,
\label{rad2b}
\end{eqnarray}
where $V_{2b}(r)$ is the two-body interaction (assumed to be central),
$\langle \rangle_\Omega$ indicates integration over the angles only,
and $E_{tot}$ is the total relative two-body energy. The
energy $E$ of the two-body system will be obtained after subtraction of
the harmonic oscillator energy, i.e.  $E=E_{tot}-\hbar \omega/2$.

An important point is that the squeezing potential (\ref{trap}) is not
central, and therefore the orbital angular momentum quantum number,
$\ell$, is not conserved.  In other words, the trap potential is not
diagonal in $\ell\ell'$.  In particular we have:
\begin{equation}
\langle Y_{\ell m}|V_{trap}(r,\theta)|Y_{\ell' m'}\rangle_\Omega = \frac{1}{2}\mu\omega^2 r^2
\langle Y_{\ell m}|\cos^2\theta|Y_{\ell' m'}\rangle_\Omega,
\end{equation}
and
\begin{eqnarray}
\lefteqn{
\langle Y_{\ell m}|\cos^2\theta|Y_{\ell' m'}\rangle_\Omega= }\nonumber \\ &&
\delta_{mm'}(-1)^m\sum_L (2L+1) \sqrt{2\ell+1} \sqrt{2\ell'+1} \times \nonumber \\ && \times
\left(
        \begin{array}{ccc}
        1  &  1  &  L  \\
        0  &  0  &  0
        \end{array}
\right)^2
\left(
        \begin{array}{ccc}
        \ell  &  L  &  \ell'  \\
          0   &  0  &  0
        \end{array}
\right)
\left(
        \begin{array}{ccc}
        \ell  &  L  &  \ell'  \\
         -m   &  0  &  m
        \end{array}
\right),
\label{cos2}
\end{eqnarray}
where the brackets are $3j$-symbols, 
$L$ can then obviously only take the values $L=0$ and $L=2$, and
therefore $\ell+\ell'$ has to be an even number (so, the parity is
well-defined). Note that the angular momentum projection $m$ actually
remains as a good quantum number throughout the transition to 2D.
Therefore, the value taken for $m$ in Eq.(\ref{cos2}) characterizes
the solution in 2D obtained after infinite squeezing.

In the 3D-limit ($V_{trap}=0$) the partial waves decouple, and the 3D wave function
has, of course,  a well-defined orbital angular momentum. In the calculations reported in this work the 3D wave function
will be assumed to have $\ell=0$, which therefore means that $m=0$. 

Note that if we take $m_\omega=\mu$, and we make use of Eqs.(\ref{eq1}) and (\ref{trap}), the coupled equations
(\ref{rad2b}) can be written as:
\begin{eqnarray}
\lefteqn{ 
\left[ \frac{\partial^2}{\partial r_b^2}-\frac{\ell(\ell+1)}{r_b^2}-
2 V_{2b}^b(r) + 2 E_{tot}^b
\right] u_\ell} \label{rad2b2}\\ && \hspace*{5mm}
-\frac{r_b^2}{(b^b_{ho})^4} 
    \sum_{\ell' m'}\langle Y_{\ell m}|\cos^2\theta|Y_{\ell' m'}\rangle_\Omega
                               u_{\ell'}=0,
\nonumber
\end{eqnarray}
where, taking $b$ as some convenient length unit, we have defined $r_b=r/b$, $b_{ho}^b=b_{ho}/b$, 
$V_{2b}^b=V_{2b}/(\hbar^2/\mu b^2)$, and $E_{tot}^b=E_{tot}/(\hbar^2/\mu b^2)$. In other words, when taking
$b$ and $\hbar^2/\mu b^2$ as length and energy units, respectively, the two-body radial equation,
Eq.(\ref{rad2b2}), is independent of the reduced mass of the system.

\subsection{2D~$\rightarrow$~1D}
\label{2d1d}

The confinement from two to one dimensions can be made in a similar way as done in the
previous subsection for the 3D~$\rightarrow$~2D case, that is, solving the two-dimensional two-body problem
with an external squeezing potential along one direction (that we choose along the $y$-axis).

As before, when working in the center of mass frame, the external potential will be given by
\begin{equation}
V_{trap}=\frac{1}{2}\mu\omega^2 r^2 \sin^2\varphi,
\label{trap2}
\end{equation} 
where, as in Eq.(\ref{trap}), the radial coordinate $r$ is the relative distance between the two particles, but
where now the polar angle $\varphi$ is such that $x=r\cos\varphi$ and $y=r\sin\varphi$. In
this way, after infinite squeezing, the particles are allowed to move along the $x$-axis only.

In 2D the partial wave expansion of the wave function, analogous to Eq.(\ref{eq3}), is given by:
 \begin{equation}
 \Psi(\bm{r}) =\sum_m \frac{u_m(r)}{\sqrt{r}} Y_m(\varphi),
 \label{eq11}
 \end{equation}  
 where the angular functions
 \begin{equation}
 Y_m( \varphi) =\frac{1}{\sqrt{2\pi}} e^{im\varphi}
 \label{eq13}
 \end{equation}
 are the eigenfunctions of the 2D angular momentum operator $-i\hbar\partial/\partial \varphi$, whose eigenvalue
 $\hbar m$ can take positive and negative values, where $m$ is an integer.
 
 Using the expansion in Eq.(\ref{eq11}), the 2D radial Schr\"{o}dinger equation then reads:
 \begin{eqnarray}
 \lefteqn{
 \left[
 \frac{\partial^2}{\partial r^2} +  \frac{\frac{1}{4}-m^2}{r^2}-\frac{2\mu}{\hbar^2}V_{2b}(r)+\frac{2\mu E_{tot}}{\hbar^2}
 \right]u_m }
 \nonumber
 \\ & &
 -\frac{2\mu}{\hbar^2}\sum_{m'} \langle Y_m|V_{trap}(r,\varphi)|Y_{m'} \rangle_\Omega u_{m'}=0,
\label{sch2D}
 \end{eqnarray}
 which is equivalent to Eq.(\ref{rad2b}).   Using Eq.(\ref{eq13}), it is not difficult to see that
 \begin{equation}
\langle Y_m|V_{trap}(r,\varphi)|Y_{m'}\rangle_\Omega = \frac{1}{2}\mu\omega^2 r^2
\langle Y_m|\sin^2\varphi|Y_{m'}\rangle_\Omega,
\label{eq15}
\end{equation}
and
 \begin{equation}
 \langle Y_m|\sin^2 \varphi|Y_{m'}\rangle_\Omega = 
 \frac{1}{2}\delta_{m,m'}-\frac{1}{4}\delta_{m,m'\pm2},
 \label{eq16}
 \end{equation}
 which implies that, as in the 3D~$\rightarrow$~2D case, the squeezing potential is again mixing different
 angular momentum quantum numbers.

The procedure shown up to here is completely analogous to the one described in the
previous subsection for 3D~$\rightarrow$~2D squeezing. However, in this case the starting
point is the 2D Schr\"{o}dinger equation Eq.(\ref{sch2D}), which shows the important feature that
for $s$-waves ($m=0$) the ``centrifugal'' barrier is actually attractive, and it takes
the very particular form of $-1/4r^2$. This barrier
happens to be precisely the critical potential giving rise to the ``falling to the
center'' or Thomas effect \cite{lan77}. As a consequence, the numerical resolution of the
differential equation (\ref{sch2D}) can encounter difficulties associated to this pathological 
behavior.

To overcome this numerical problem, it is more convenient to face the 2D~$\rightarrow$~1D squeezing
problem solving the Schr\"{o}dinger equation directly in Cartesian coordinates:
\begin{equation}
\left[ -\frac{\hbar^2}{2\mu}\left( \frac{\partial^2}{\partial x^2}+
\frac{\partial^2}{\partial y^2} \right) +V(x,y)+V_{trap}(y)-E_{tot}\right] \Psi=0.
\label{schxy}
\end{equation}
This can be easily made after expanding the two-body relative wave function, 
$\Psi$, in an appropriate basis set  where the Hamiltonian can be diagonalized. A possible choice for
the  basis could be $\{ |\psi_{n_x}(x) \psi_{n_y}(y)\rangle\}$,
with $\psi_n$ being the harmonic oscillator eigenfunctions.
Needless to say, solving either Eq.(\ref{sch2D}) or (\ref{schxy}) is completely
equivalent, although in (\ref{schxy}) the problematic attractive barrier is not
present, and the dependence on the angular momentum $m$ also disappears. 

Again, as discussed in Eq.(\ref{rad2b2}), after taking $m_\omega=\mu$ in Eq.(\ref{eq1}), and $b$ and
$\hbar^2/\mu b^2$ as length and energy units, respectively, Eq.(\ref{sch2D}) (or (\ref{schxy}))
becomes $\mu$-independent.

\subsection{3D~$\rightarrow$~1D}
\label{3to1}

If the confinement procedures 3D~$\rightarrow$~2D and 2D~$\rightarrow$~1D described in sections \ref{3a} and \ref{2d1d} are performed consecutively, we are then 
obviously performing a 3D~$\rightarrow$~1D squeezing, but going through an intermediate 2D geometry. However, this is not really necessary, since it is 
always possible to squeeze the system in two directions simultaneously, which will lead to the same 1D space but without going through the 2D configuration.

Let us consider that the particles, in principle moving in a 3D space, are confined be means of two external harmonic oscillator potentials acting on the $x$ and
the $y$ directions simultaneously. The trap potential felt by each of the particles is then:
\begin{equation}
V_{trap}^{(i)}=\frac{1}{2}m_i \omega_x^2x_i^2+\frac{1}{2}m_i \omega_y^2 y_i^2,
\label{eq18}
\end{equation}
where $x_i$ and $y_i$ are the $x$ and $y$ coordinates of particle $i$, and $\omega_x$ and $\omega_y$ are the harmonic oscillator frequencies of each of the
two external potentials. These frequencies determine the independent squeezing
on each of the directions, and of course the infinitely many possible values
of the $\omega_x/\omega_y$-ratio determine the infinitely many possible ways of squeezing from 3D into 1D.

Let us here consider the simplest case in which $\omega_x=\omega_y=\omega$. After using spherical coordinates, we trivially get that:
\begin{equation}
V^{(i)}_{trap}=\frac{1}{2}m_i\omega^2 r_i^2 \sin^2\theta_i=\frac{1}{2} \frac{m_i \hbar^2}{m_\omega^2 b_{ho}^4} r_i^2 \sin^2\theta_i,
\label{eq19}
\end{equation}
which, as one could expect, is identical to Eq.(\ref{eq2}) but replacing  $\cos\theta_i$ by $\sin\theta_i$. This simply means that the vector coordinate $\bm{r}_i$ is not
projected on the $z$-axis, but on the $xy$-plane.

Therefore, the discussion below Eq.(\ref{eq2}) still holds here, but replacing $\cos\theta$ by $\sin\theta$ all over, which leads again to Eq.(\ref{rad2b})
but where now
\begin{equation}
\langle Y_{\ell m}|V_{trap}(r,\theta)|Y_{\ell' m'}\rangle_\Omega = \frac{1}{2}\mu\omega^2 r^2
\langle Y_{\ell m}|\sin^2\theta|Y_{\ell' m'}\rangle_\Omega,
\label{eq20}
\end{equation}
with
\begin{equation}
\langle Y_{\ell m}|\sin^2\theta|Y_{\ell'm'}\rangle_\Omega=
\delta_{\ell\ell'}\delta_{mm'}-\langle Y_{\ell m}|\cos^2\theta|Y_{\ell'm'}\rangle_\Omega,
\label{eq21}
\end{equation}
and where the last matrix element is given by Eq.(\ref{cos2}).

Since in this case we have included two harmonic oscillator potentials, the energy provided by them,
still assuming $\omega_x=\omega_y=\omega$, will be $E_{ho}=\hbar\omega$,  and therefore
$E=E_{tot}-\hbar\omega$.

\section{Two-body systems in $d$-dimensions}
\label{sec3}

An alternative to the continuous squeezing of the particles by means of external potentials
can be to solve the two-body problem in $d$ dimensions, where $d$ can take any value 
within the initial and final dimensions ($3\geq d \geq 2$,  $2\geq d \geq 1$, or $3\geq d \geq 1$). 
The idea is that, given a squeezed system by means of an external potential with squeezing
parameter $b_{ho}$, it is then possible to associate this particular squeezing parameter to some
specific non-integer value of the dimension, such that the properties of the system
can be obtained by solving the, in general simpler, $d$-dimensional problem. 
The basic properties of two-body systems in $d$ dimensions are
described in Appendix C of Ref.\cite{nie01}. For this reason, in this section we just
collect the key equations relevant for the work presented here.

\subsection{Theoretical formulation}

Let us consider a two-body system where the relative coordinate between the two constituents
is given by $\bm{r}$. In principle the components of the vector $\bm{r}$ in a $d$-dimensional
space will be given by the $d$ Cartesian coordinates $(r_1,r_2,\cdots,r_d)$. As it is well known, 
when dealing with central potentials, it is however much more convenient to use the set of generalized 
spherical coordinates, which
contain just one radial coordinate $r=\sqrt{r_1^2+r_2^2+\cdots+r_d^2}$ and $d-1$ angles (for instance
the polar and azimuthal angles when $d=3$). In this way, the two-body wave function
can be expanded in terms of the generalized $d$-dimensional spherical harmonics, which depend
on the $d-1$ angles (see \cite{nie01} for details):
\begin{equation}
\Psi_d(\bm{r})=\frac{1}{r^{\frac{d-1}{2}}}\sum_\nu R^{(d)}_\nu(r)Y_\nu(\Omega_d),
\label{wf}
\end{equation}
where $\nu$ represents the summation over all the required quantum numbers, $\Omega_d$ collects
the $d-1$ angular coordinates, and where $\int Y_\nu^*Y_{\nu'} d\Omega_d=\delta_{\nu\nu'}$.

Of course, what written above makes full sense provided that $d$ takes integer values. However, when
$d$ is not an integer, which obviously will happen when changing the dimension 
continuously from the initial down to the final dimension, the meaning of the 
$d-1$ angles in Eq.(\ref{wf}) is not obvious. Nevertheless, in this work we are 
not having this problem, since we are considering relative $s$-waves only, which implies
that  the wave function (\ref{wf}) is angle independent, and it can actually be written as:
\begin{equation}
\Psi_d(\bm{r})=\frac{1}{r^{(d-1)/2}} R_d(r) Y_d,
\label{wf2}
\end{equation}
where the constant $s$-wave spherical harmonic, $Y_d$, can be obtained simply by keeping in mind that
in $d$ dimensions we have that \cite{hay01}: 
\begin{equation}
\int d\Omega_d=\frac{2\pi^{d/2}}{\Gamma\left(\frac{d}{2}\right)}, 
\end{equation}
which, making use of the fact that $Y_d^* Y_d\int d\Omega_d=1$, immediately leads to:
\begin{equation}
Y_d=\left[ \frac{\Gamma\left(\frac{d}{2}\right)}{2\pi^{d/2}}\right]^{1/2}.
\label{Yd}
\end{equation}

Finally, the radial wave function, $R_d(r)$, in Eq.(\ref{wf2}) is obtained as the solution 
of the $d$-dimensional radial Schr\"{o}dinger equation, which for $s$-waves reads \cite{nie01}:
\begin{equation}
\left[\frac{\partial^2}{\partial r^2} -\frac{\frac{1}{4}(d-3)(d-1)}{r^2}
             -\frac{2\mu}{\hbar^2}(V_{2b}(r)-E)\right]R_d(r)=0,
\label{2bdd}
\end{equation}
where, since the only potential entering is just the two-body interaction, the energy $E$ is the
true two-body  relative energy.

It is important to note that
if we write $d=2+x$ with $-1\leq x \leq 1$, the barrier in Eq.(\ref{2bdd}) takes the form $(x^2-1)/4r^2$, which
indicates that the equation to be solved is the same for $d=2-x$ and $d=2+x$. This might suggest that the bound
state solutions of Eq.(\ref{2bdd}) should be symmetric around $d$=2. For instance, for $d$=1 and $d$=3 the barrier disappears,
 and one could expect the same solution in the two cases. However, as discussed below, this is not really like this.

Note that the centrifugal barrier in Eq.(\ref{2bdd}) can also be written in the usual way, $\ell^*(\ell^*+1)/r^2$, simply 
by defining $\ell^*=(d-3)/2$. This means that Eq.(\ref{2bdd}) is formally identical to the usual radial two-body 
Schr\"{o}dinger equation with angular momentum $\ell^*$. Therefore, as it is well known, Eq.(\ref{2bdd}) has in general two
possible solutions, each of them associated to a different short-distance behavior:
\begin{eqnarray}
&R_d^{(1)}(kr)\stackrel{kr \rightarrow 0}{\longrightarrow} kr j_{\ell^*}(kr) \stackrel{kr \rightarrow 0}{\longrightarrow} r^{\ell^*+1}=r^{\frac{d-1}{2}}, \label{rad1} \\
&R_d^{(2)}(kr)\stackrel{kr \rightarrow 0}{\longrightarrow} kr \eta_{\ell^*}(kr) \stackrel{kr \rightarrow 0}{\longrightarrow} r^{-\ell^*}=r^{\frac{3-d}{2}} \label{rad2},
\end{eqnarray}
where $j_{\ell^*}$ and $\eta_{\ell^*}$ are the regular and irregular spherical Bessel functions, respectively, and $k=\sqrt{2\mu |E|}/\hbar$. 
Within the dimension range $1\leq d \leq 3$, 
these two radial solutions go to zero for $kr\rightarrow 0$, except for $d=1$ and $d=3$, where
one of the solutions goes to a constant value. 

However, as shown in Eq.(\ref{wf2}), the full radial wave function is actually $R_d(r)/r^{(d-1)/2}$.
After dividing by the phase space factor we immediately see that the solution in Eq.(\ref{rad1}) leads to a constant value of the full radial wave function
at $r=0$ no matter the dimension $d$. Thus, the solution (\ref{rad1}) is valid in the full dimension range $1\leq d\leq 3$.
However, after dividing by the phase space factor,  the radial solution obtained from Eq.(\ref{rad2}) behaves
at short distances as $r^{2-d}$. This means that this solution is regular at the origin only for $d<2$, whereas it has to be disregarded
for $d>2$. For $d>2$ only the solution (\ref{rad1}), whose short-distance behavior is determined by the regular Bessel function, is physically acceptable.

\begin{figure}[t]
\centering
\includegraphics[width=0.8\linewidth]{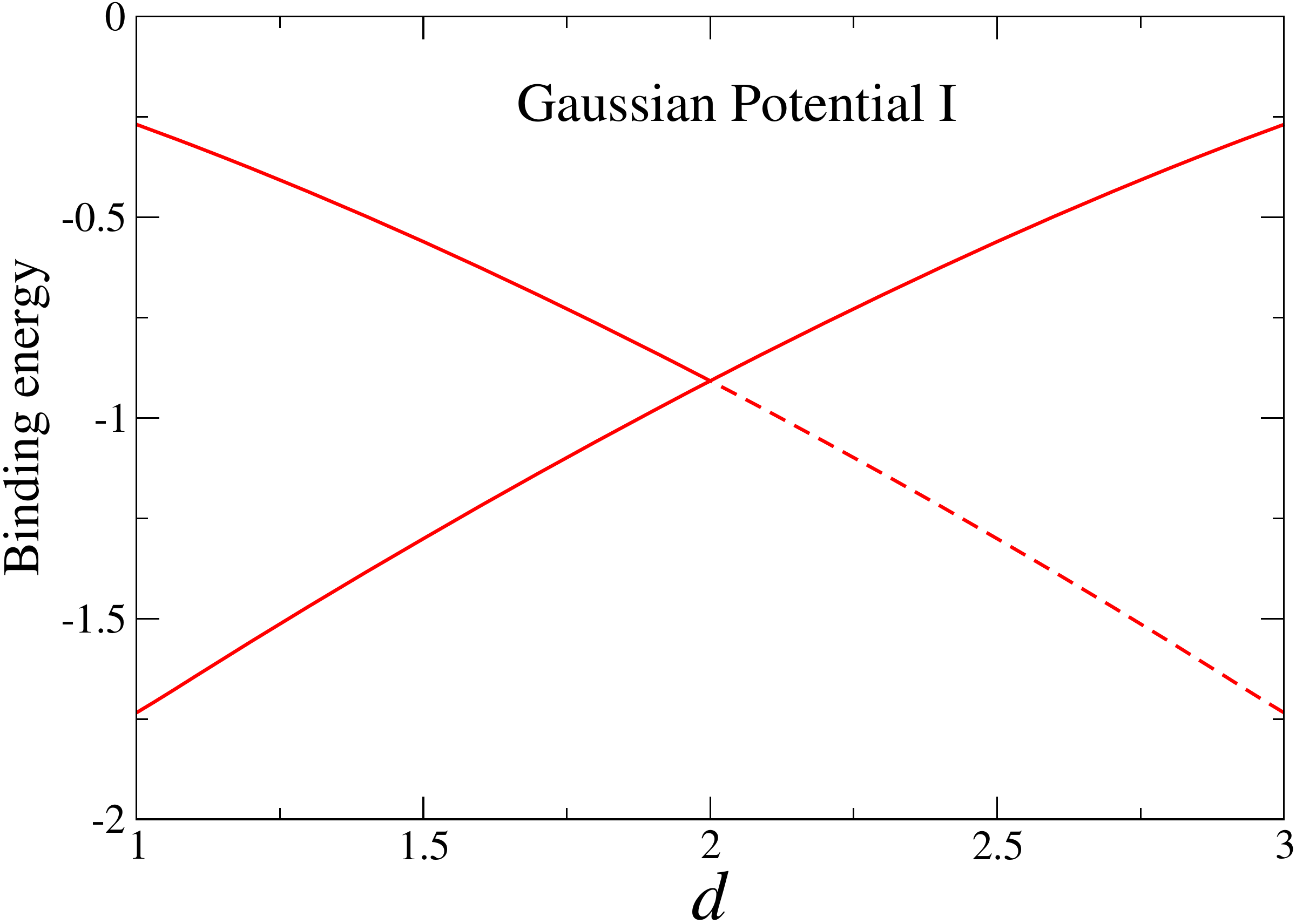}
\caption{Binding energies obtained from Eq.(\ref{2bdd}) as a function of $d$. The two-body potential used is the Gaussian potential indicated as potential I in
Table~\ref{tab1}. The dashed line corresponds to the states to be discarded due to the divergence of the solution at the origin.
   }
\label{fig1}     
\end{figure}

Therefore the apparent symmetry of Eq.(\ref{2bdd}) around $d=2$ is not real. The two-body problem has two solutions with physical meaning for $d<2$, and only one for $d>2$.
Furthermore, the solution to be disregarded for $d>2$ corresponds to a larger binding, which implies that the ground state of the system will be more bound
for $d=1$ than for $d=3$. This is illustrated in Fig.~\ref{fig1}, where we show the two computed binding energies arising from Eq.(\ref{2bdd})
as a function of the dimension $d$.

Although not relevant at this stage, let us mention for completeness that the two-body potential used 
in the calculation is the one that later on will be called Gaussian potential I.
The branch indicated in Fig.~\ref{fig1} by the dashed curve for $d>2$ is the one obtained after imposing to the solution
the short-distance behavior  (\ref{rad2}), which, as discussed above, can not be accepted as a physical solution due to the divergence of the full radial wave function
at the origin. As seen also in the figure, for $d=2$ the two solutions merge into a single one, as expected due to the fact that the short-distance 
behavior (\ref{rad1}) and (\ref{rad2}) is the same in this case.

\subsection{Interpretation of the wave function}
\label{sec2a}

Once the $d$-dimensional radial wave function, $R_d(r)$, has been obtained, it is possible to compute  different
observables in $d$ dimensions, as for instance the root-mean-square radius, which is given by the simple expression
\begin{equation}
r_d^2=\int_{0}^{\infty} r^2 \left|R_d(r)\right|^2 dr.
\label{rd}
\end{equation}

However, the reliability of a direct use of a non-integer $d$-dimensional wave function, $\Psi_d$, to compute 
a given observable, that unavoidably is measured in a three- or two-dimensional space, is not obvious.
It could actually look more convenient to use instead the wave function obtained with
an external squeezing potential,  which, although very often more
difficult to compute, is, in the strict sense, a three- or two-dimensional wave function.

In order to exploit the simplicity in the calculation of the $\Psi_d$ wave function, it is
necessary to obtain a procedure to translate the non-integer $d$-dimensional wave function
into the ordinary three- or two-dimensional space. 

To do so, let us start by noticing that a
squeezing of the system by an external field acting along one (for the 3D~$\rightarrow$~2D and 
2D~$\rightarrow$~1D cases) or two (for the 3D~$\rightarrow$~1D case) directions,
necessarily proceeds through deformed structures in the initial dimension.
It appears then necessary to provide a reinterpretation 
of the total spherical wave function in $d$ dimensions, 
$\Psi_d$, as corresponding to a deformed three-dimensional system (for the 3D~$\rightarrow$~2D 
and 3D~$\rightarrow$~1D cases), or a deformed two-dimensional system (for the 2D~$\rightarrow$~1D
case). The simplest way to account for this is to deform the radial coordinate $r$
along the squeezed direction(s). In this way, if we consider the usual Cartesian coordinates
$\{x,y,z\}$ for an initial 3D system, or the $\{x,y\}$ coordinates for an initial 2D system,
we can interpret $\Psi_d$ as an ordinary three- or two-dimensional wave function, but
where the radial argument, $r$, is replaced by $\tilde{r}$, which is defined as:
\begin{equation}
r\rightarrow \tilde{r} \equiv \sqrt{x^2+y^2+(z/s)^2} \equiv \sqrt{r_\perp^2+(z/s)^2} \;,
\label{scl1}
\end{equation}\begin{equation}
r\rightarrow \tilde{r} \equiv \sqrt{(x^2+y^2)/s^2+z^2} \equiv \sqrt{(r_\perp/s)^2+z^2} \;,
\label{scl2}
\end{equation}
\begin{equation}
r\rightarrow \tilde{r} \equiv \sqrt{x^2+(y/s)^2}\;,
\label{scl3}
\end{equation}
for the 3D~$\rightarrow$~2D, 3D~$\rightarrow$~1D and 2D~$\rightarrow$~1D cases,
respectively. 

In the expressions above, $s$ is a scale parameter, assumed to be independent of the value of
the squeezed coordinate, and which, in principle, lies within
the range $0\leq s \leq 1$.  For $s=1$, the relative radial
coordinate $\tilde{r}$ is the usual one in spherical or polar coordinates, and the system is not 
deformed. For $s=0$, only $z=0$ in Eq.(\ref{scl1}),
$r_\perp=0$ in Eq.(\ref{scl2}), and $y=0$ in Eq.(\ref{scl3}) are possible, otherwise $\tilde{r}=\infty$ no
matter the value of the non-squeezed coordinate, 
and, since we are dealing with bound systems, $\Psi_d(\tilde{r})=\Psi_d(\infty)=0$. Therefore, the $s=0$ situation
corresponds to a completely squeezed system into two dimensions
when (\ref{scl1}) is used (3D~$\rightarrow$~2D), or into one dimension when (\ref{scl2}) or 
(\ref{scl3}) is used (3D~$\rightarrow$~1D or 2D~$\rightarrow$~1D). 

It is important to note that after the transformation of the radial coordinate as defined in 
Eqs.(\ref{scl1}) to (\ref{scl3}), the $d$-dimensional  wave function, $\Psi_d$, has to be normalized in 
the new three-dimensional space for the 3D~$\rightarrow$~2D and 3D~$\rightarrow$~1D cases:
\begin{equation}
2\pi\int r_\perp dr_\perp dz |\tilde{\Psi}_d(r_{\perp},z,s)|^2=1,
\label{norm1}
\end{equation}
or in the new two-dimensional space for the 2D~$\rightarrow$~1D case:
\begin{equation}
\int dx dy |\tilde{\Psi}_d(x,y,s)|^2=1,
\label{norm2}
\end{equation}
where $\tilde{\Psi}_d \propto \Psi_d$ denotes the normalized wave function.

With this interpretation, where $\tilde{\Psi}_d$ is a wave function in the ordinary 3D (or 2D) space, one can obtain 
the expectation value of any observable ${\cal F}(\bm{r})$ in the usual way, that is:
\begin{equation}
\langle {\cal F}(\bm{r})\rangle_s=\int r_\perp dr_\perp dz d\varphi {\cal F}(\bm{r})|\tilde{\Psi}_d(r_\perp,z,s)|^2
\label{exp1}
\end{equation}
or
\begin{equation}
\langle {\cal F}(\bm{r})\rangle_s=\int dx dy {\cal F}(\bm{r})|\tilde{\Psi}_d(x,y,s)|^2.
\label{exp2}
\end{equation}

The remaining point here is how to determine the value of the scale
parameter, $s$, corresponding to a specific squeezing produced by an
external field with a given oscillator parameter, $b_{ho}$.  The
interpretation of $\Psi_d$ as an ordinary (deformed) wave function
using a constant $s$ is very tempting. We believe this must be correct
to leading order. If the Schr\"{o}dinger equations are approximately solved by 
variation using single-gaussian solutions we can directly identify the matching scale 
factor $s$. For instance, in case of squeezing along the $z$-direction, the
two single-gaussian solutions will have the form 
$R_{ext}\propto e^{-r_\perp^2/2b^2-z^2/2b_z^2}$ when the
external field is used, and $R_d\propto e^{-r^2/2b_d^2}$ after the $d$-calculation.
If, following Eq.(\ref{scl1}), we interpret $R_d$ as a function of $\tilde{r}$, we 
easily get that $R_{ext}$ and $R_d$ are the same if $b=b_d$ and $s=b_z/b_d$.  
One way to improve is to allow $s$ to be a function of 
the squeezed coordinate. The assumption is mostly that this dependence on the squeezed coordinate 
is rather smooth, in such a way that the function can be safely expanded around some constant 
average value, which therefore is the leading term. 

In any case, the scale parameter $s$ has to be a function
of the non-integer dimension parameter, $d$, or equivalently, the
squeezing length parameter, $b_{ho}$.  The value of the average scale
parameter, $s$, then has to be obtained by comparison of the wave
functions from a full external field, $\Psi_{b_{ho}}(r)$, with
$r^2=r_\perp^2+z^2$ (when squeezing from 3D) or $r^2=x^2+y^2$ (when
squeezing from 2D), and the normalized $d$-dimensional wave function,
$\tilde\Psi_d(\tilde{r})$, defined above.

Being more precise, we define the overlaps:
\begin{equation}
{\cal O}_{3D}(s)=2\pi \int r_\perp dr_\perp dz \tilde{\Psi}_d(r_\perp,z,s) \Psi_{b_{ho}}(r) \; ,
\label{ov1}
\end{equation}
and
\begin{equation}
{\cal O}_{2D}(s)=\int dx dy \tilde{\Psi}_d(x,y,s) \Psi_{b_{ho}}(r) \; ,
\label{ov2}
\end{equation}
which are valid for initial 3D and 2D spaces, respectively. The scale parameter, $s$,
is then determined such that the overlap (\ref{ov1}) for the 
3D~$\rightarrow$~2D and 3D~$\rightarrow$~1D cases, or
the overlap (\ref{ov2}) for the 2D~$\rightarrow$~1D case, is maximum.

In Appendix \ref{app} we show,  Eqs.(\ref{rat1}), (\ref{rat2}) and (\ref{rat3}), that
the scale factor, $s$,   is actually given by
\begin{equation}
s=\left( \frac{\langle z^2 \rangle_s }{\langle z^2 \rangle_{s=1} } \right)^{1/2};
s=\left( \frac{\langle r_\perp^2 \rangle_s }{\langle r_\perp^2 \rangle_{s=1} } \right)^{1/2};
 s=\left( \frac{\langle y^2 \rangle_s }{\langle y^2 \rangle_{s=1} } \right)^{1/2},
 \label{eq39}
\end{equation}
for the 3D~$\rightarrow$~2D,  3D~$\rightarrow$~1D, and  2D~$\rightarrow$~1D
squeezing cases, respectively, where $\langle \rangle_s$ are expectation values
as defined in Eqs.(\ref{exp1}) and (\ref{exp2}). This result says that $s$ is nothing but 
the ratio  between the expectation value of the squeezing coordinate for that value
of $s$, and the one obtained without deforming the wave function ($s=1$). 
These expressions make evident that the scale parameter, $s$, is a measure of the deformation
along the squeezing direction(s).

\section{Large squeezing regime}
\label{app2}

The large-squeezing region corresponds to very small values of the
oscillator parameter, $b_{ho}$, which implies that the squeezing potential
dominates over the two-body interaction along the squeezing
direction(s).  As a consequence, the root-mean-square value of a given squeezed coordinate $u$ is, for large squeezing,
essentially given by the one corresponding to the harmonic oscillator potential. In particular this implies that,
for a squeezing process $d_i$D~$\rightarrow$~$d_f$D we can write, in a compact way:
\begin{equation}
u_{rms} = \langle u^2\rangle^{1/2} \stackrel{b_{ho} \rightarrow 0}{\longrightarrow} \sqrt{\frac{d_i-d_f}{2}} b_{ho},
\label{eqb1}
\end{equation}
where $u$ can be $z$, $r_\perp$, or $y$ for the 3D~$\rightarrow$~2D, 3D~$\rightarrow$~1D, and 2D~$\rightarrow$~1D cases,
respectively, and $d_i$ and $d_f$ indicate the initial and final dimension.
 
 Let us now focus on the radius, $r_d$, defined in Eq.(\ref{rd}). To simplify ideas, let us consider first the case of a 3D~$\rightarrow$~2D squeezing along the $z$-coordinate,
 and write $r_d$ as 
 \begin{equation}
 r_d^2=\langle r^2 \rangle=\langle r_\perp^2\rangle + \langle z^2\rangle.
 \label{eqb4}
 \end{equation}
 
 Since the squeezing takes place along the $z$ direction, the expectation value $\langle z^2\rangle$ is the one feeling the squeezing effect, whereas 
 $\langle r_\perp^2\rangle$ to leading order does not. Following  Eq.(\ref{eq39}) we can then write:
 \begin{equation}
 r_d^2=\langle r_\perp^2\rangle + s^2 \langle z^2\rangle_{s=1},
 \end{equation}
 where $\langle z^2\rangle_{s=1}$ is the expectation value of $\langle z^2\rangle$ when $\Psi_d$ is interpreted as non-deformed standard 3D wave function. 
 Furthermore,  for a spherical 3D wave function we know that
 $ \langle x^2\rangle=\langle y^2\rangle=\langle z^2\rangle$, which means that $\langle r_\perp^2\rangle=\langle x^2\rangle+\langle y^2\rangle=2\langle z^2\rangle_{s=1}$,
 and we can then write:
 \begin{equation}
 r_d=\sqrt{2+s^2} \langle z^2\rangle_{s=1}^{1/2}=\frac{\sqrt{2+s^2}}{s} \langle z^2\rangle^{1/2},
 \label{eqb6}
 \end{equation}
  for 3D~$\rightarrow$~2D, where Eq.(\ref{eq39}) has again been used.
 
 The same argument is valid for the 3D~$\rightarrow$~1D case, except for the fact that now in Eq.(\ref{eqb4}) the squeezing is felt by $\langle r_\perp^2\rangle$
 and not by $\langle z^2\rangle$. Therefore, using  again Eq.(\ref{eq39}), we can write $\langle r_\perp^2\rangle=s^2 \langle r_\perp^2\rangle_{s=1}$, and
 $\langle z^2\rangle=\langle r_\perp^2\rangle_{s=1}/2$. In this way we get for 3D~$\rightarrow$~1D:
\begin{equation}
 r_d=\sqrt{ \frac{1+2 s^2}{2}} \langle r_\perp^2\rangle_{s=1}^{1/2}=\sqrt{\frac{1+2s^2}{2s^2}} \langle r_\perp^2\rangle^{1/2}.
 \label{eqb7}
 \end{equation}
 
 Finally, similar arguments for the 2D~$\rightarrow$~1D case, where $r_d^2=\langle r^2 \rangle=\langle x^2\rangle + \langle y^2\rangle$, assuming the squeezing
 along $y$, lead to:
 \begin{equation}
 r_d=\sqrt{1+s^2} \langle y^2\rangle_{s=1}^{1/2}=\frac{\sqrt{1+s^2}}{s} \langle y^2\rangle^{1/2}.
 \label{eqb8}
 \end{equation}
 
 From Eqs.~(\ref{eqb6}), (\ref{eqb7}), and (\ref{eqb8}) it is easy to obtain that, to leading order:
 \begin{equation}
 s \approx  \sqrt{\frac{d_f}{d_i-d_f} } \frac{u_{rms}/r_d }{\sqrt{1-\left( u_{rms}/r_d\right)^2}},
 \label{eqb46}
 \end{equation}
 where $u_{rms}\equiv\langle u^2\rangle^{1/2}$, and $u$ represents either $z$, $r_\perp$, or $y$, depending on what squeezing process we are dealing with.

 At this point it is easy to replace in Eq.(\ref{eqb46})  the large squeezing behavior
 of $\langle z^2\rangle$, $\langle r_\perp^2\rangle$, and $\langle y^2\rangle$ given in Eq.(\ref{eqb1}), and obtain the following
 expression for the scale parameter, $s$, in case of large squeezing:
 \begin{equation}
 s \stackrel{b_{ho}\rightarrow 0 }{\longrightarrow} \left[ \frac{d_f\left(\frac{b_{ho}}{r_d} \right)^2}{2-(d_i-d_f)\left( \frac{b_{ho}}{r_d}\right)^2}\right]^{1/2},
 \label{eqb12}
 \end{equation}
where $d_i$ and $d_f$ again denote the initial and final dimension.

\begin{table}[t]
\begin{tabular}{c|ccc}
                   & Pot. I & Pot. II  &   Pot. III    \\ \hline
    \multicolumn{4}{c}{ } \\ 
    $S_g$                     &    $-2.71$       &   $-1.43$                      &   $-1.38$                                   \\ \hline
    $E_{3D}$    &     $-0.269$    &  $-1.651\cdot10^{-3}$ &   $-3.144\cdot10^{-4}$            \\ 
    $E_{2D}$    &     $-0.908$    &   $-0.269$                     &   $-0.249$                         \\
    $E_{1D}$    &     $-1.734$    &   $-0.771$                     &   $-0.736$                         \\ \hline
    $a_{3D}$                &     2.033          &   18.122                        &    40.598                            \\
    $a_{2D}$                &      1.103         &     1.883                        &      1.942                             \\
    $a_{1D}$                &        0.063      &               0.899            &         0.928                             \\ \hline
     $r_{3D}$               &       1.508       &    12.823                     &    28.710                                   \\  
     $r_{2D}$               &       0.926       &      1.398                     &       1.439                            \\      
     $r_{1D}$              &       0.572       &    0.747                      &        0.759                          \\    \hline
      \multicolumn{4}{c}{ } \\
    $S_m$                     &    $1.294$       &   $0.474$                      &   $0.434$                                \\ \hline
    $E_{3D}$    &     $-0.189$    &  $-1.875\cdot10^{-3}$ &   $-3.325\cdot10^{-4}$           \\ 
    $E_{2D}$    &     $-0.450$    &   $-7.394\cdot10^{-2}$&   $-6.088\cdot10^{-2}$          \\
    $E_{1D}$    &     $-0.811$    &   $-0.228$                     &   $-0.203$                            \\ \hline
    $a_{3D}$                &     2.033          &   18.122                        &    40.598                            \\
    $a_{2D}$                &     $\sim 10^{-8}$  &     3.536                 &      3.872                              \\
    $a_{1D}$                &        14.0      &               0.986            &         1.243                         \\ \hline
     $r_{3D}$               &       2.235       &    12.870                   &    28.741                                  \\  
     $r_{2D}$               &       1.458      &      2.739                    &       2.947                             \\      
     $r_{1D}$               &       0.875       &    1.366                     &        1.428                             \\    \hline
\end{tabular}
\caption{Strengths corresponding to the three Gaussian, $S_g$, and Morse, $S_m$, potentials used. For each of them we give the 
$s$-wave two-body binding energies of the ground state in three, two, and one dimensions ($E_{3D}$, $E_{2D}$, and  $E_{1D}$), 
the corresponding $s$-wave scattering lengths $a_{3D}$, $a_{2D}$ and $a_{1D}$, and the root-mean-square radii $r_{3D}$,
$r_{2D}$, and $r_{1D}$. All the energies are given in units of $\hbar^2/\mu b^2$ and the lengths in units of $b$, where $b$ is
the range of either the Gaussian or the Morse potential.}
\label{tab1}
\end{table}

\section{Potentials}
\label{sec5}

Let us start the numerical illustration by first specifying the chosen potentials along with a few of their
characteristic properties.  In the following subsections we continue
to present results of the two methods described formally in
Sections~\ref{sec2} and \ref{sec3}.

\subsection{General properties}

To allow general, and hopefully universal conclusions, we shall use
two different radial shapes for the two-body potential: A Gaussian
potential, $V_{2b}(r)=S_g e^{-r^2/b^2}$, and a Morse-like potential,
$V_{2b}(r)=S_m(e^{-2r/b}-2e^{-r/b})$. The range of the interaction,
$b$, will be taken as the corresponding (in principle different) length unit.
Therefore, as discussed in section \ref{3a}, taking $m_\omega=\mu$ in
Eq.(\ref{eq1}) and $\hbar^2/\mu b^2$ as energy unit, the
Schr\"{o}dinger equation is independent of the
reduced mass.

For each of the potential shapes, three different interactions will be
considered, potentials I, II, and III. The potential parameters for
each of them are chosen such that the 3D scattering length, $a_{3D}$,
is the same for both, the Gaussian and the Morse shapes.  The
$a_{3D}$ values are in all the cases positive (therefore holding a 3D
bound state), and change from comparable to the potential range, $b$,
for potential I, to about 20 times $b$ for potential II, up to a value
of about 40 times $b$ for potential III.

The details of the potentials are given in Table~\ref{tab1} for the
employed Gaussian and Morse shapes, where the lengths are in units of
$b$, and the energies, including the strengths, $S_g$ and $S_m$, are in
units of $\hbar^2/\mu b^2$.  The characterizing $s$-wave scattering
lengths, $a_{3D}$, $a_{2D}$, and $a_{1D}$, are obtained in
three-dimensional, two-dimensional, and one-dimensional calculations,
respectively.  The two-dimensional scattering length $a_{2D}$ is defined
as given in Eq.(C.6) in Ref.\cite{nie01}. Furthermore, we give in Table~\ref{tab1} the
corresponding $s$-wave ground state binding energies, $E_{3D}$,
$E_{2D}$, and $E_{1D}$ together with their root-mean-square radii,
$r_{3D}$, $r_{2D}$, and $r_{1D}$.

All the potentials given in Table~\ref{tab1} give rise to only
one bound state, i.e., the ground state. The only exception is Potential I
with Morse shape in one dimension. The two-body system described by 
this potential has a weakly bound excited state, such that 
the large value of the scattering length $a_{1D}$ is in this case
related to the appearance of this second state, whose energy 
can be approximated by some constant divided by $a_{1D}^2$
(Eq.(C.9) in Ref.\cite{nie01}).

\subsection{External harmonic oscillator potential}

Let us start with the case of confinement by means of an external (harmonic oscillator) potential. 
The procedure is as described in section~\ref{sec2}, where it is shown how the trap potential, which
is not central, mixes different values of the relative orbital angular momentum. This is made evident
in Eqs.(\ref{cos2}), (\ref{eq16}), and (\ref{eq21}) for the three different squeezing processes
(3D~$\rightarrow$~2D, 2D~$\rightarrow$~1D, 3D~$\rightarrow$~1D) considered in this work.
Therefore, it is not difficult to foresee that
the stronger the squeezing, the larger the number of terms in the expansions (\ref{eq3}) and (\ref{eq11})
required to get convergence.  In fact, for no squeezing, the orbital angular momentum is a good quantum number
and only one term in the expansion enters.

\begin{figure}[t]
\centering
\includegraphics[width=0.95\linewidth]{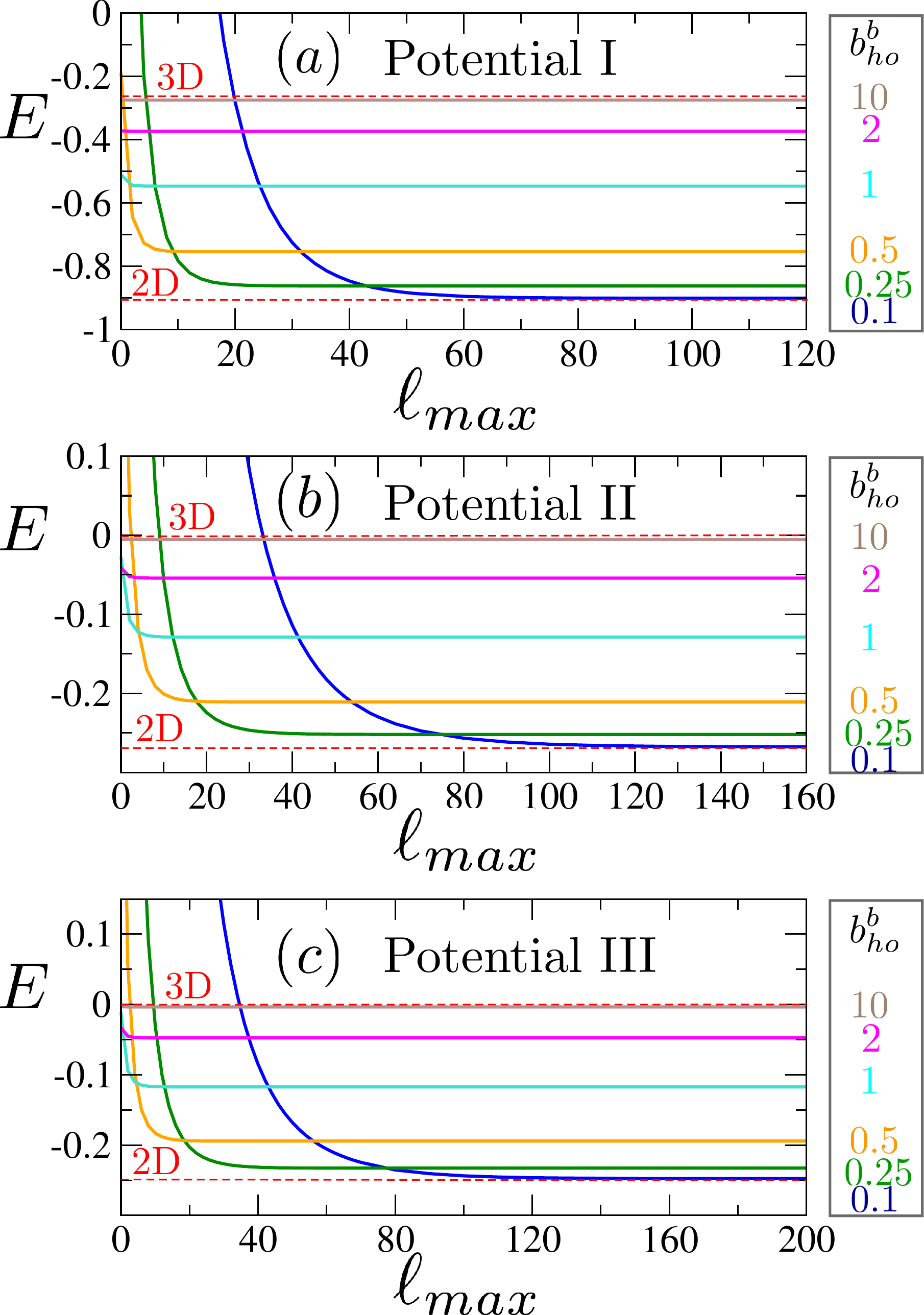}
\caption{Two-body energies (after subtracting the harmonic oscillator energy), in units
of $\hbar^2/\mu b^2$, for the Gaussian potentials in a 3D~$\rightarrow$~2D squeezing,
as a function of the $\ell_{max}$ value included in the expansion (\ref{eq3}). The results
for $b_{ho}^b=b_{ho}/b=$ 0.1, 0.25, 0.5, 1, 2, and 10 are shown. Panels $(a)$, $(b)$, and $(c)$ refer
to potentials I, II, and III, respectively. On each panel, the upper and lower horizontal dashed 
lines indicate the two-body energies in the 3D and 2D cases, respectively, as indicated by
the ``3D'' and ``2D'' labels.
   }
\label{fig2}     
\end{figure}

In order to illustrate the pattern of convergence, let us consider the 3D~$\rightarrow$~2D case and 
call $\ell_{max}$ the maximum value of $\ell$ included in  the expansion (\ref{eq3}).
In Fig.~\ref{fig2}a we show the convergence of the two-body energy as a function of $\ell_{max}$ for
the Gaussian potential I (Table~\ref{tab1}), and for different values of $b_{ho}^b=b_{ho}/b$ in the squeezing
potential. The energy shown, $E$, is the two-body energy obtained after subtracting the
harmonic oscillator energy, i.e., $E=E_{tot}-E_{ho}$. The horizontal dashed lines are the two-body energies
obtained after a 3D and a 2D calculation, respectively, which are given in Table~\ref{tab1}.

As expected, for small values of the oscillator parameter we recover the computed 2D-energy, whereas
for large values of $b_{ho}^b$ the 3D-energy is approached. We can also see that the smaller $b_{ho}^b$, 
the larger the $\ell_{max}$ value needed to get convergence. Partial waves with $\ell$-values up to
around 80 are at least needed for $b_{ho}^b=0.1$, for which we get a converged energy of $-0.901$, pretty
close to the value of $-0.908$ obtained in a 2D-calculation. For large values of $b_{ho}^b$ the convergence 
is obviously much faster. For $b_{ho}^b=10$ we obtain an energy of $-0.274$, not far from the
value of $-0.269$ corresponding to the 3D calculation. This result is already obtained including the $\ell=0$ 
component only.

In Fig.~\ref{fig2}b we show the same as in Fig.~\ref{fig2}a for the Gaussian potential II. 
The general features are the same as before, although there are some remarkable
differences arising from the fact that now the scattering length is about 9 times
bigger. First, for $b_{ho}^b=10$ we obtain a converged energy of 
$-5.43\cdot 10^{-3}$, which, although in the figure seems to be very close to
3D energy, it differs by more than a factor of three ($E_{3D}=-1.65\cdot 10^{-3}$). 
To get a better agreement with the 3D-energy, $b_{ho}^b$ values of a few times  the 
3D scattering length are needed (as in fact observed for potential I).
The second important difference is that convergence is now slower than before, and higher 
values of $\ell_{max}$ are needed to get convergence for small oscillator lengths.

These facts are more emphasized when using the Gaussian potential III, whose corresponding curves
are shown in Fig.~\ref{fig2}c. In this case we have obtained for $b_{ho}^b=10$ 
a converged energy of $-3.41\cdot 10^{-3}$, about an order of magnitude more bound than
the 3D energy ($-3.14\cdot 10^{-4}$). For $b_{ho}^b=0.1$, an $\ell_{max}$
value of at least 140 is needed in order to get convergence. This is of course related
to the large 3D scattering length. In three dimensions the bound two-body system  is
clearly bigger than with the other two potentials, and, consequently, it starts feeling
the confinement sooner than in the other cases. 

The convergence features for 2D~$\rightarrow$~1D and 3D~$\rightarrow$~1D confinement, as well
as for the Morse potentials, are the same as the ones described in Fig.~\ref{fig2},
namely, for large values of $b_{ho}^b$ (little squeezing) a small number of partial waves are enough
to get convergence and the energy in the initial dimension, 2D or 3D, is approached, whereas for small values of  $b_{ho}^b$ 
(large squeezing) a higher number of partial waves is required, and the energy in the final dimension,
1D or 2D, is recovered. For
this reason we consider it to be unnecessary to show the corresponding figures. In any case, as discussed in 
section~\ref{2d1d}, for 2D~$\rightarrow$~1D squeezing it is more convenient to solve
directly Eq.~(\ref{schxy}), where the partial wave expansion does not enter explicitly. 

\begin{figure}[t]
\centering
\includegraphics[width=7cm]{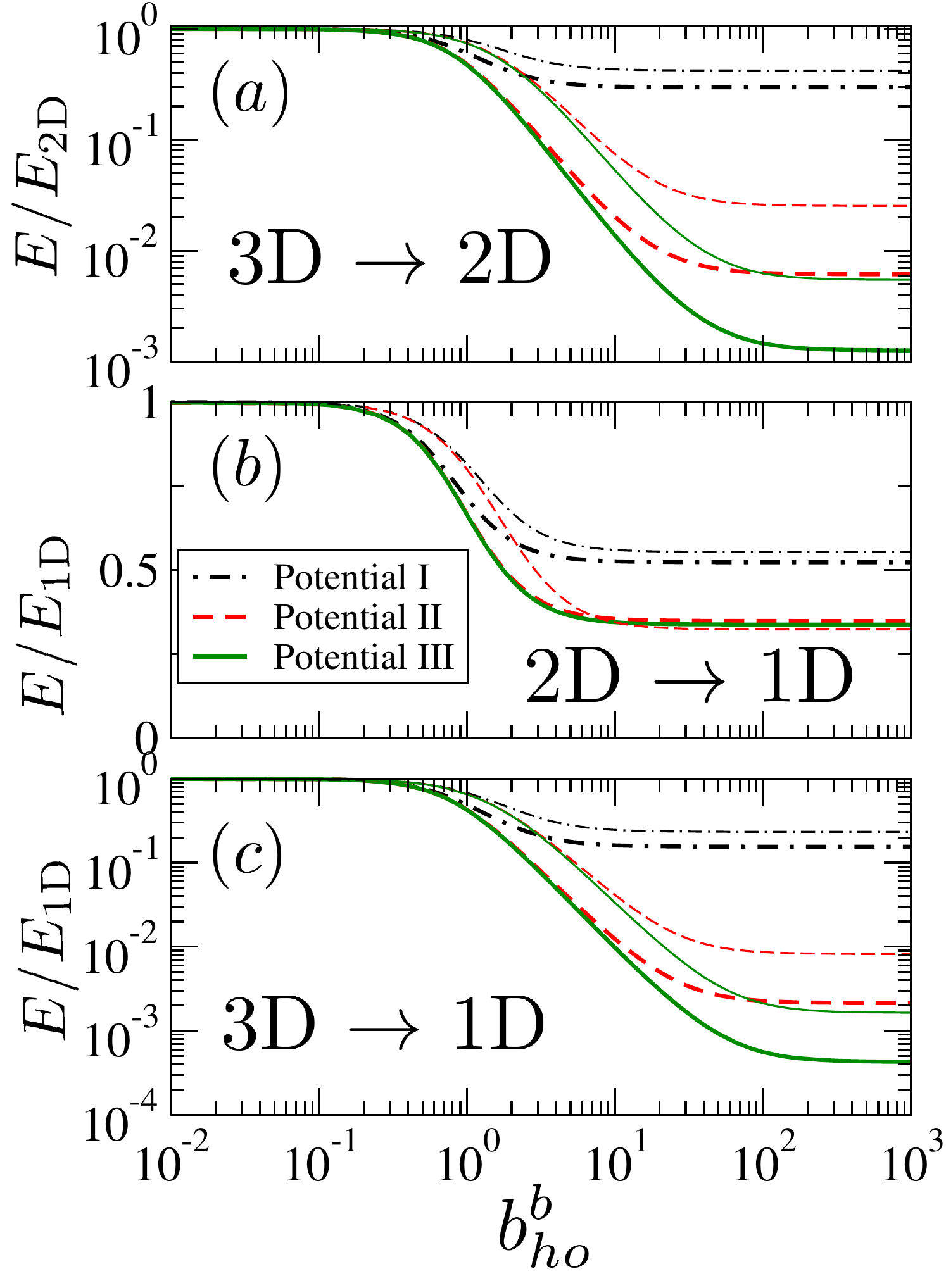}
\caption{Converged two-body energies (normalized to the energy in the final dimension) for 
the three Gaussian (thick curves) and Morse (thin curves) potentials as a function of $b_{ho}^b$. Panels (a), (b), and (c)
correspond to 3D~$\rightarrow$~2D, 2D~$\rightarrow$~1D, and  3D~$\rightarrow$~1D 
squeezing, respectively. 
}
\label{fig3}     
\end{figure}

In Figs.~\ref{fig3}a, \ref{fig3}b, and
\ref{fig3}c we show, for the 3D~$\rightarrow$~2D, 2D~$\rightarrow$~1D, and 3D~$\rightarrow$~1D 
cases, respectively, the converged values of the two-body energy $E$ for the Gaussian potentials (thick curves) and
the Morse potentials (thin curves), as a function of the oscillator parameter, $b_{ho}^b$. 
The energy is normalized to the energy in the final dimension, either $E_{2D}$ or  $E_{1D}$, given in Table~\ref{tab1}. Therefore, 
for small values of $b_{ho}^b$ all the curves go to 1. 

\subsection{Two-body energy in $\bm{d}$ dimensions}

As described in Section~\ref{sec3}, the continuous squeezing of the system from some initial dimension
$d_i$ to some final dimension $d_f$ can also be made by solving the two-body problem in $d$ dimensions, 
where $d_f\leq d \leq d_i$.
We can therefore compute the same observable as in Fig.~\ref{fig3} but solving the two-body 
Schr\"{o}dinger equation (\ref{2bdd}), where the dimension $d$ is taken
as a parameter. It is important to remember that for $d>2$ the ground state solution behaves
for $\kappa r\rightarrow 0$  as given in Eq.(\ref{rad1}), whereas for $d<2$ the ground state follows the
behavior given by Eq.(\ref{rad2}).

\begin{figure}[t]
\centering
\includegraphics[width=7cm]{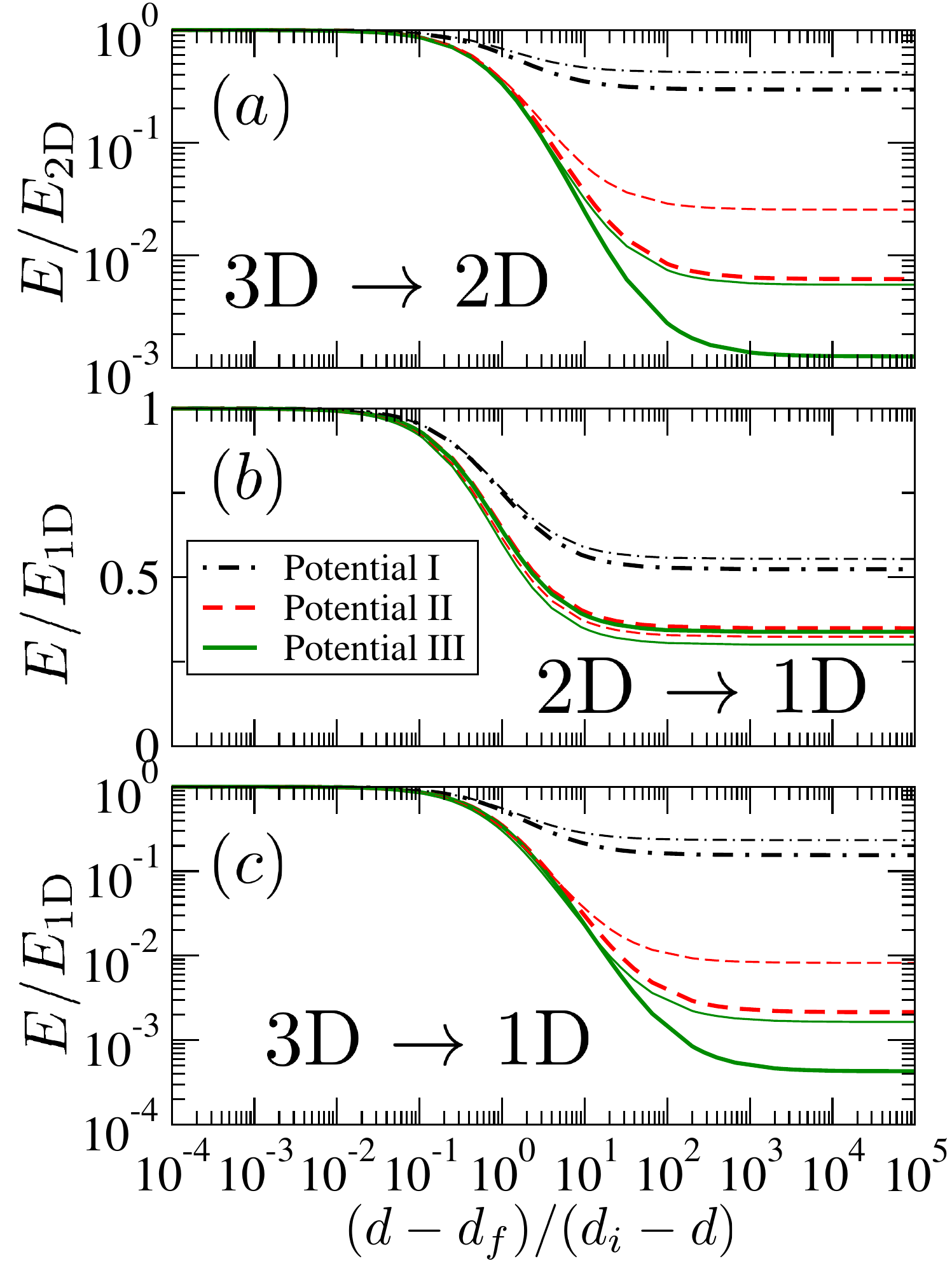}
\caption{Two-body energies (normalized to the energy in the final dimension) for 
the three Gaussian (thick curves) and the three Morse (thin curves) potentials used 
as a function of $(d-d_f)/(d_i-d)$, where $d$ is the dimension
varying continuously from the initial dimension $d_i$ to the final dimension $d_f$.
Panels (a), (b), and (c) correspond to 
3D~$\rightarrow$~2D, 2D~$\rightarrow$~1D, and  3D~$\rightarrow$~1D 
squeezing, respectively.}
\label{fig4}     
\end{figure}

The results are shown in Fig.~\ref{fig4} as a function of $(d-d_f)/(d_i-d)$ for the squeezing
cases 3D~$\rightarrow$~2D (panel (a)), 2D~$\rightarrow$~1D (panel (b)),
and  3D~$\rightarrow$~1D (panel (c)). Again, the thick and thin curves correspond to the
results with the Gaussian and Morse potentials, respectively. The choice of the 
abscissa coordinate is such that the curves can be easily compared to the ones in 
Fig.~\ref{fig3}. In fact, a simple eye inspection of both figures makes evident the existence of 
a univocal connection between $b_{ho}^b$ and $d$.

\section{Comparing the two methods}
\label{sec6}

Although the parameters used on each of the two methods, $b_{ho}$ and $d$, 
have a very different nature, they both are used to describe the same physics 
process of squeezing the system into a
lower dimensional space.  We shall first connect these parameters by
use of the energies leading from initial to final dimension in the
squeezing processes.  Then we turn to the crucial comparison of the
related wave functions which require an interpretation and a
deformation parameter, as described in Section~\ref{sec2a}.

\subsection{Relation between $b_{ho}^b$ and $d$}

\begin{figure}[t]
\centering
\includegraphics[width=7cm]{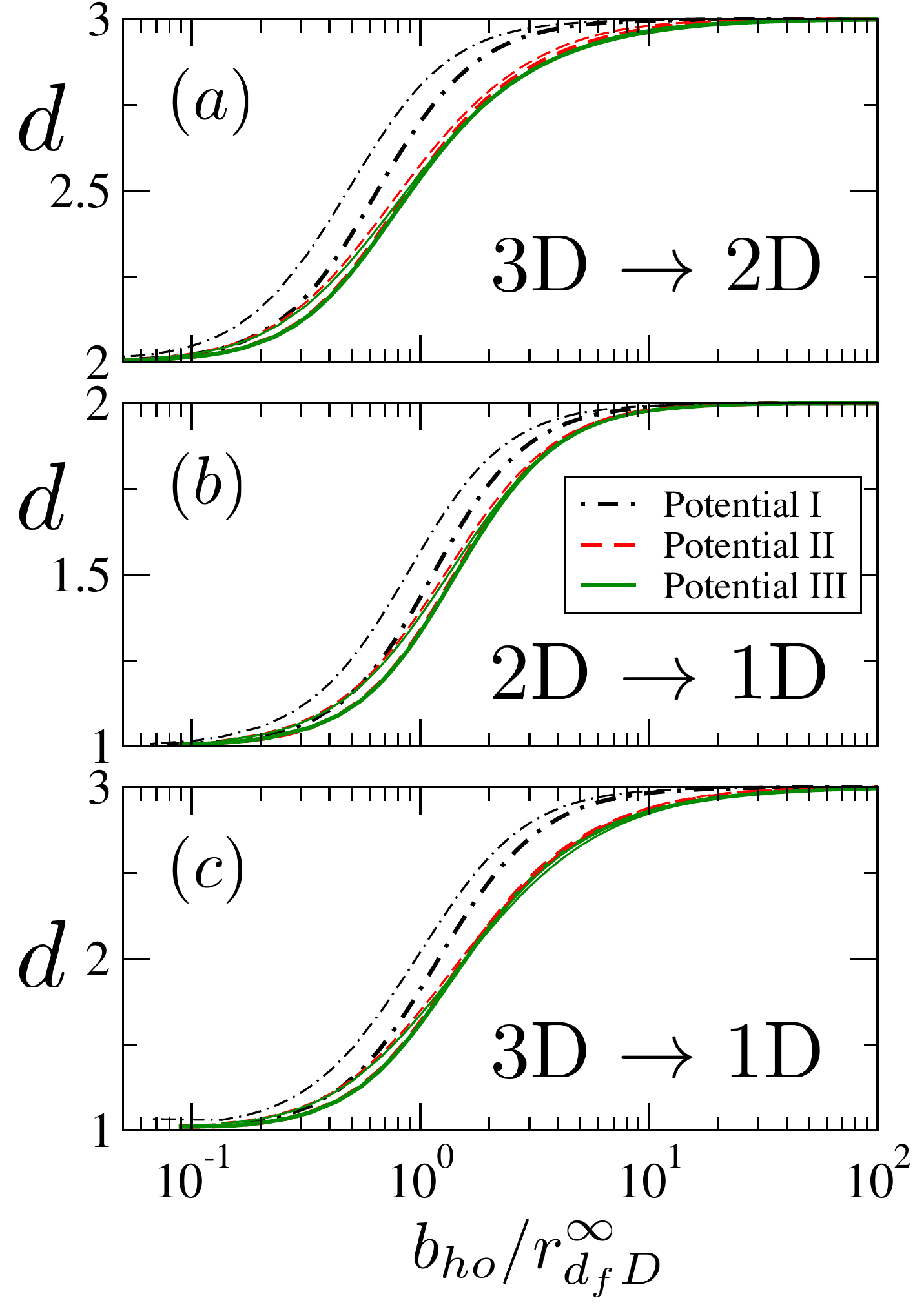}
\caption{Values of $d$ as a function of $b_{ho}/r_{d_f D}^\infty$, obtained by matching the energies
in Figs.~\ref{fig3} and Fig.~\ref{fig4}, for the potentials in Table~\ref{tab1}. The
cases of 3D~$\rightarrow$~2D, 2D~$\rightarrow$~1D and 3D~$\rightarrow$~1D squeezing are shown 
in panels (a), (b) and (c), respectively. The thick and thin-dashed curves correspond to the results 
obtained for the Gaussian and Morse potentials, respectively. The radius$r_{d_fD}^\infty$ in the root-mean-square radius in the final
dimension, $d_f$, obtained with a potential such that $a_{3D}=\infty$. }
\label{fig5}     
\end{figure}

The relation between $b_{ho}^b$ and $d$ obtained directly from Figs.~\ref{fig3} and \ref{fig4} is shown 
in Fig.~\ref{fig5} for all the potentials. Panels (a), (b), and (c) correspond to the 3D~$\rightarrow$~2D, 
2D~$\rightarrow$~1D, and 3D~$\rightarrow$~1D cases, respectively. As in the previous figures,
the thick and thin curves are, respectively, the results obtained with the Gaussian and Morse potentials.

In the figure we show $d$ as a function of $b_{ho}/r_{d_fD}^\infty$, where $r_{d_fD}^\infty$ is the root-mean-square radius 
of the bound two-body system in the final dimension obtained with an interaction such that $a_{3D}=\infty$. 
In particular $r_{1D}^\infty$ and $r_{2D}^\infty$ take the values 0.769 and 1.474, in units of $b$, for the Gaussian
shape, and 1.478 and 3.128 for the Morse shape, respectively.  
This is a way to normalize the size of the bound state in the 
final dimension, $d_f$, to the value corresponding to the potential, which, in principle, is expected to provide a universal
connection between $d$ and $b_{ho}$.  In fact, as shown in the figure, for each of the two potential shapes, 
the curves corresponding to the potentials with large scattering length, potentials II and III, are almost identical to each other. Furthermore,
the curves for these two potentials corresponding to the Gaussian (thick curves) and Morse (thin curves) potentials
are not very different. Only the cases corresponding to potential I give rise to curves clearly different to the other ones. 

This result is consistent with the idea of relating universal properties of quantum systems to the presence of  
relative s-waves and large scattering lengths. This has been established as a universal parameter describing properties 
of weakly bound states without reference to the responsible short-range attraction. For this reason,
the translation between $b_{ho}$ and $d$ shown in Fig.~\ref{fig5} for the potentials with large scattering length should be
very close to the desired universal relation between the two parameters. 

\begin{figure}[t]
\centering
\includegraphics[width=7cm]{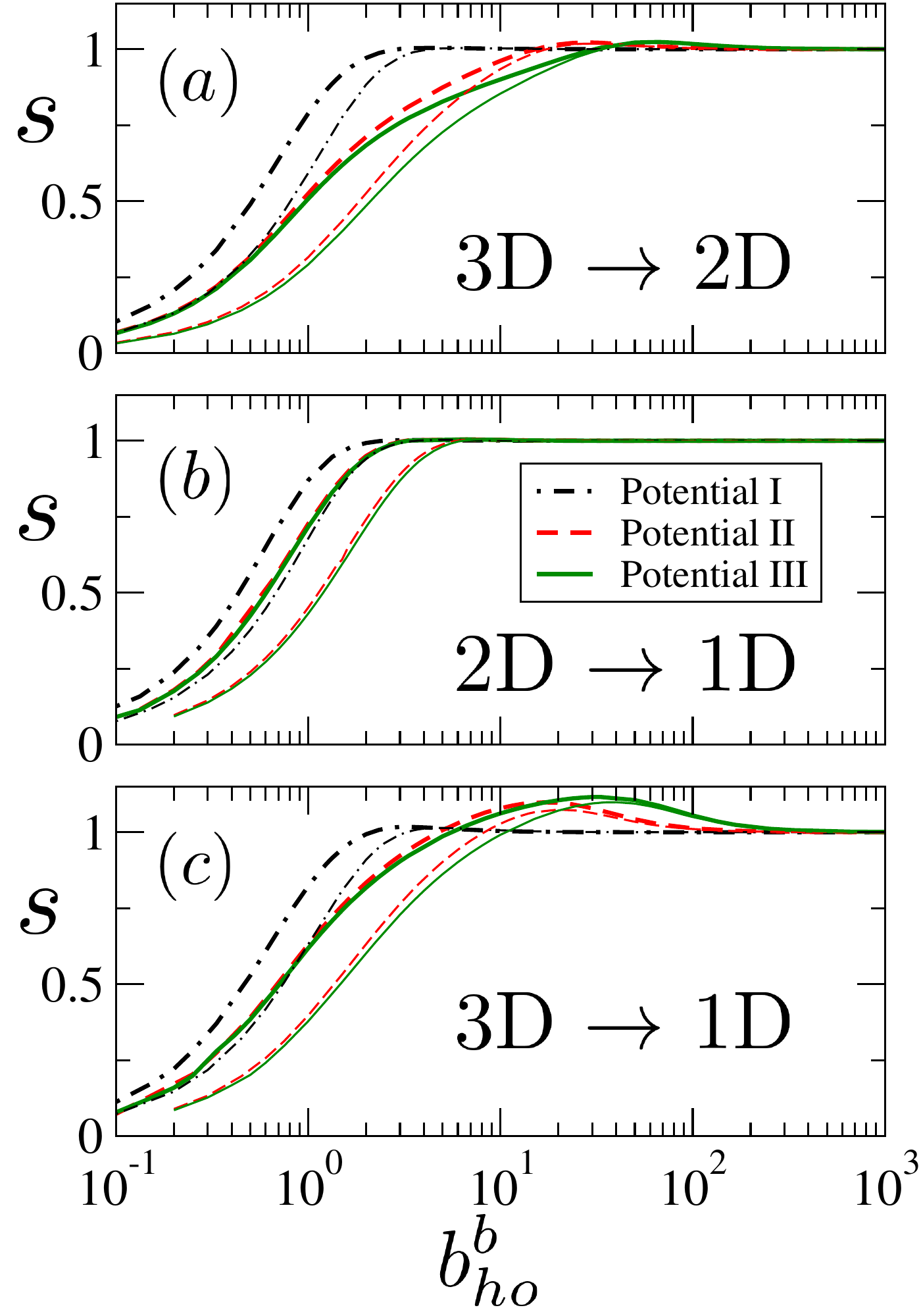}
\caption{The scale parameter $s$ as a function of $b_{ho}^b$ for the potentials in 
Table~\ref{tab1} and the three squeezing scenarios considered in this work.}
\label{fig6}
\end{figure}

\subsection{Scale parameter}

As discussed in subsection~\ref{sec2a}, the wave function in $d$ dimensions can
be interpreted as an ordinary wave function in three dimensions (in the 3D~$\rightarrow$~2D or 3D~$\rightarrow$~1D cases)
or in two dimensions (in the 2D~$\rightarrow$~1D case), but with a deformation
along the squeezing direction, as shown in Eqs.~(\ref{scl1}), (\ref{scl2}), and (\ref{scl3}).
The value of the scale parameter, $s$, is obtained as the one maximizing the overlap, ${\cal O}_{3D}$,
in Eq.(\ref{ov1}) for the 3D~$\rightarrow$~2D or 3D~$\rightarrow$~1D cases, or the overlap, 
${\cal O}_{2D}$, in Eq.(\ref{ov2}) for the 2D~$\rightarrow$~1D case. These overlaps, which are
functions of the
scale parameter, $s$, are just the overlap between the wave function obtained with the
external squeezing potential, $\Psi_{b_{ho}}$, and the renormalized wave function, $\tilde{\Psi}_d$, obtained in $d$ dimensions, where $b_{ho}$ and $d$ are related as
shown in Fig.~\ref{fig5}.

The results obtained for the scale parameter are shown in Fig.~\ref{fig6}
for the three squeezing processes and the usual three potentials for both 
the Gaussian (thick curves) and Morse (thin curves) shapes. In all the cases
the maximized overlap value is very close to 1. In fact, in the most unfavorable 
computed case ($b_{ho}^b$=0.1), the overlap value is, for all the cases, at least
0.98. As expected, 
a large squeezing ($b_{ho}^b\rightarrow 0$) implies a small value of $s$ ($s\rightarrow 0$),
whereas a small squeezing (large $b_{ho}^b$) corresponds to $s\rightarrow 1$. In fact,
for $b_{ho}=\infty$, Eqs.(\ref{rad2b}) and (\ref{sch2D}) are identical to Eq.(\ref{2bdd})
for $d=3$ and $d=2$ respectively. This means that the wave functions, 
$\Psi_{b_{ho}}$ and $\Psi_d$, are identical, and the corresponding overlaps,
${\cal O}_{3D}$ or ${\cal O}_{2D}$, are trivially maximized and equal to 1 for $s=1$.

Another feature observed in Fig.~\ref{fig6} is that when the
squeezing begins, for relatively large values of $b_{ho}^b$, the scale
parameter, $s$, can be bigger than 1. This is especially true for
potentials II and III in the 3D~$\rightarrow$~1D squeezing. This fact
indicates that in this region ($d$ very close to the initial
dimension) the interpretation of the $d$-wave function as
the three-dimensional wave function, $\tilde{\Psi}_d(\tilde{r})$, gives rise to a state with the particles a
bit too confined along the squeezing direction, in such a way that
maximization of the overlap (\ref{ov1}) or (\ref{ov2}) requires a
small release of the confinement by means of a scale factor bigger
than 1. This is very likely a consequence of using a constant scale
parameter, or equivalently, that the perpendicular and squeezing
directions are not completely decoupled for these short-range
potentials.

The differences between the curves shown in Fig.~\ref{fig6} 
are related to the size of the two-body system in the initial dimension.
In panels (a) and (c), potential III describes a two-body system in 3D clearly bigger
than the other potentials (see Table~\ref{tab1}), and therefore the curve corresponding 
to this potential is the first one feeling the squeezing, i.e., it is the first one
for which $s$ deviates from 1 when the squeezing parameter, $b_{ho}^b$ decreases. 
For the same reason the second potential feeling the squeezing is potential II, and for 
potential I the deviation from $s = 1$ starts for even smaller values of $b_{ho}^b$. For the 2D~$\rightarrow$~1D case, Fig.~\ref{fig6}b, the curves
corresponding to potentials II and III are very similar, since these
two potentials describe systems with a very similar size in 2D (see
Table~\ref{tab1}).

For the same reason there is
a clear dependence on the potential shape. In general, given a squeezing parameter, the
root-mean-square radius is clearly bigger with the Morse potential than in the Gaussian case,
as seen for instance in Table~\ref{tab1} with the $r_{2D}$ and $r_{1D}$ values. This
fact implies that, for a given $b_{ho}^b$, the scale parameter in case of using the Morse
potential is clearly smaller than when the Gaussian potential is used.

\begin{figure}[t]
\centering
\includegraphics[width=7cm]{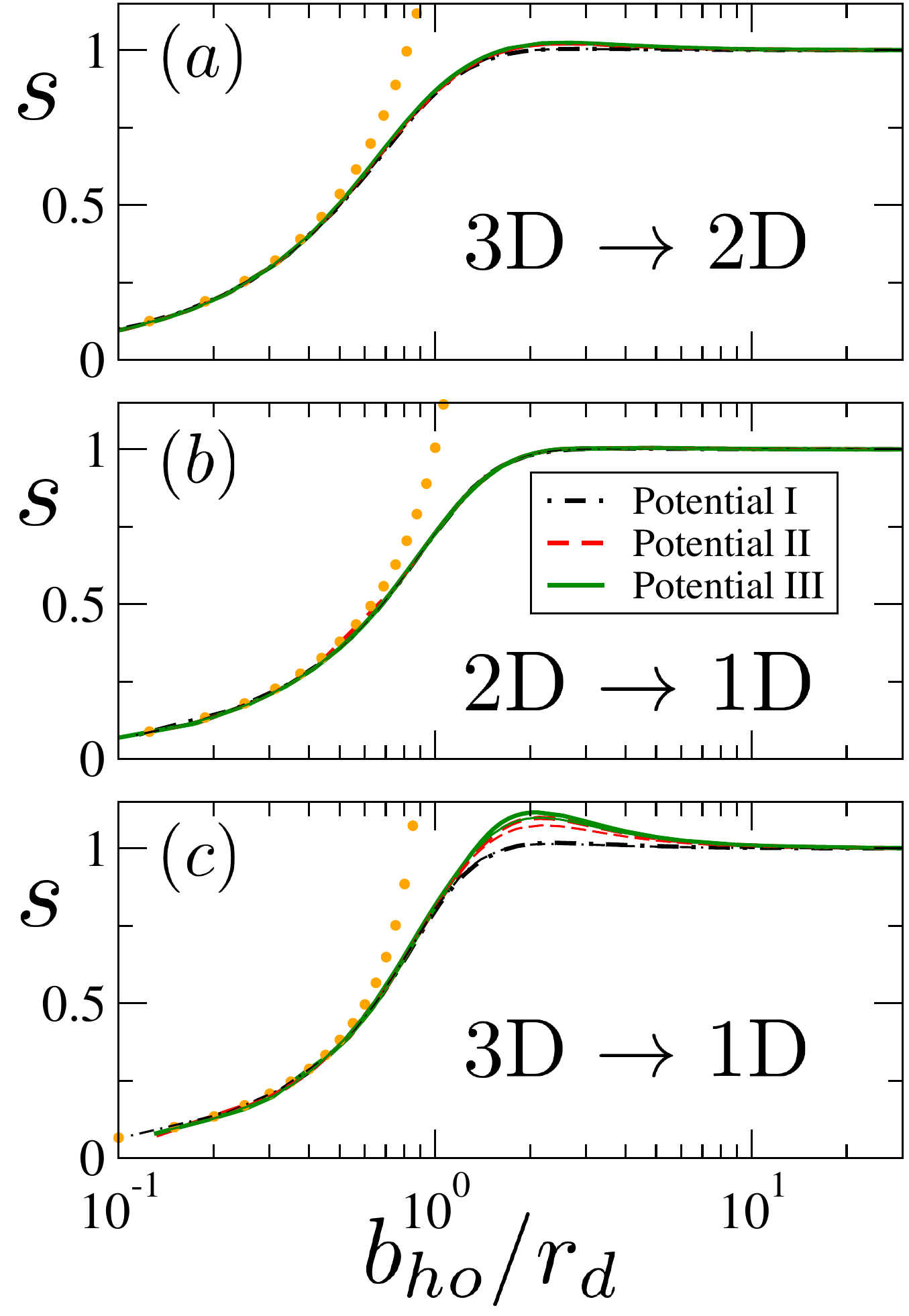}
\caption{The scale parameter $s$ as a function of $b_{ho}/r_d$, see Eq.(\ref{rd}), for the potentials in 
Table~\ref{tab1} and the three squeezing scenarios considered in this work. Except for panel (c) in the vicinity of $s=1$, 
the curves corresponding to the Gaussian (thick curves) and Morse (thin curves) potentials are, to a large extent, indistinguishable.
The circles show the analytical expression (\ref{eqb12}) valid for large squeezing.}
\label{fig7}
\end{figure}

 A simple way to account for these size effects is to plot the scale parameter, $s$, as
a function of $b_{ho}/r_d$ , where $r_d$ is the root-mean-square radius of the system for dimension $d$ 
as given in Eq.(\ref{rd}). This is shown in Fig.~\ref{fig7}, where we can see that for all the three
potentials and the Gaussian and Morse shapes, the curves collapse into a single universal curve.
The only discrepancy appears in panel (c) in the region where the squeezing begins to
produce some effect, where the bump shown by potentials II and III is not observed in the case
of potential I. This also happens, although to a much smaller extent, in the  3D~$\rightarrow$~2D case
shown in panel (a). We also show in Fig.~\ref{fig7} (dots)  the analytical expression given in Eq.(\ref{eqb12}),
which gives the relation between $s$ and $b_{ho}/r_d$ in the case of large squeezing, i.e., in the case
of small $b_{ho}$ values. As we can see, the analytical expression can be used for values of 
$b_{ho}/r_d \lesssim 0.5$.

\begin{figure}[t]
\centering
\includegraphics[width=9cm]{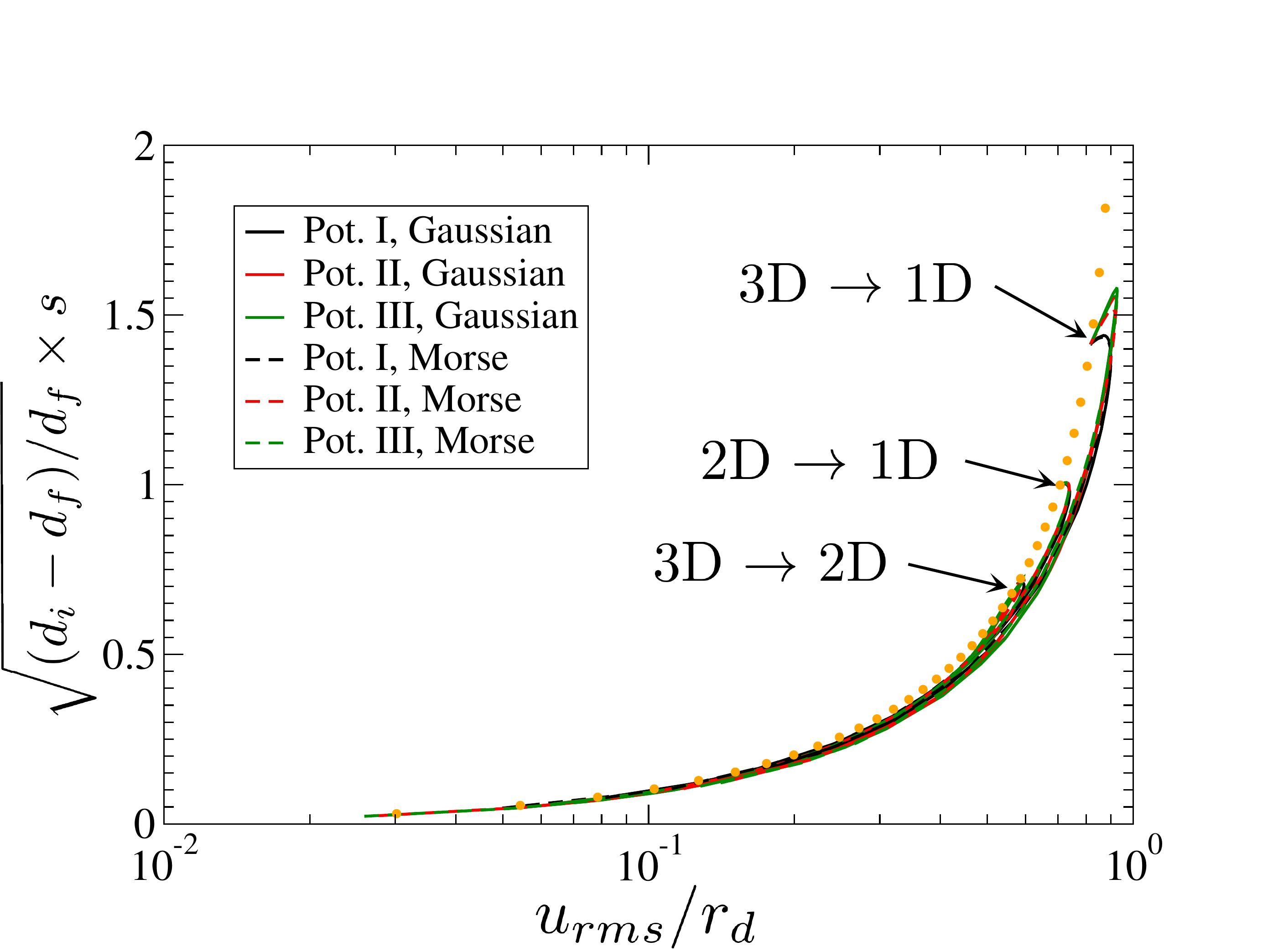}
\caption{The scale parameter $s$ ($\times \sqrt{(d_i-d_f)/d_f}$), as a function of $u_{rms}/r_d$ for the potentials in Table~\ref{tab1}, where $u=z$, $u=y$, and $u=r_\perp$
for 3D~$\rightarrow$~2D,  2D~$\rightarrow$~1D, and 3D~$\rightarrow$~1D, respectively. 
The circles show the analytical expression (\ref{eqb46}) valid for large squeezing.}
\label{fig8}
\end{figure}

It is also interesting to show the scale parameter, $s$, as a function of $u_{rms}/r_d$, where $u_{rms}=\langle u^2 \rangle^{1/2}$,
and $u$ corresponds
to $z$, $y$, or $r_\perp$ depending on what squeezing process we are dealing with, 3D~$\rightarrow$~2D, 
2D~$\rightarrow$~1D, or 3D~$\rightarrow$~1D. Such relation should be governed by the expression given in Eq.(\ref{eqb46}). 
The result is shown in Fig.~\ref{fig8}, where
the scale parameter ($\times \sqrt{(d_i-d_f)/d_f}$) is plotted as a function of $u_{rms}/r_d$. As we can see, all the curves
for all the squeezing cases are very similar to each other. The main difference appears in the low squeezing region. In
fact, in the case of no squeezing the value of $u_{rms}/r_d$ is different for each case, $1/\sqrt{3}$, $1/\sqrt{2}$,
or $\sqrt{2/3}$, as indicated by the arrows in the figure. The dotted curve is the analytical form in Eq.(\ref{eqb46}),
which for $u_{rms}/r_d \lesssim 0.4$ reproduces quite well the computed curves. 

\begin{figure}[t]
\centering
\includegraphics[width=8cm]{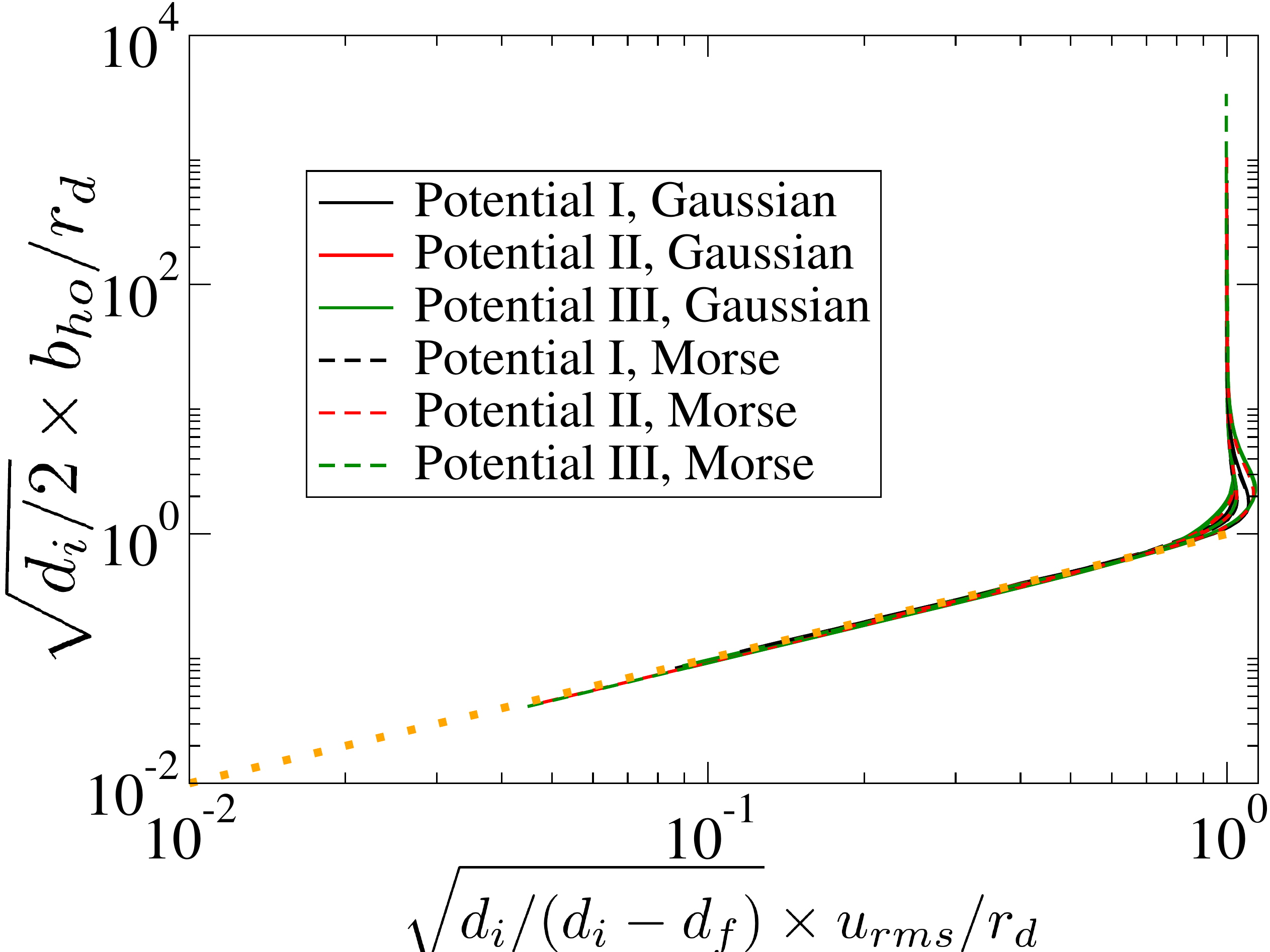}
\caption{Value of $b_{ho}/r_d$ ($\times \sqrt{d_i/2}$) as a function of $u_{rms}/r_d$ ($\times \sqrt{d_i/(d_i-d_f)}$)
for the potentials in Table~\ref{tab1}, where $u=z$, $u=y$, and $u=r_\perp$
for 3D~$\rightarrow$~2D,  2D~$\rightarrow$~1D, and 3D~$\rightarrow$~1D, respectively. 
The dotted line is obtained from Eq.(\ref{eqb1}).}
\label{fig9}
\end{figure}

Finally, it is clear from Figs.~\ref{fig7} and \ref{fig8} that it is also possible to relate $b_{ho}/r_d$ and $u_{rms}/r_d$, simply
by connecting the values of these two quantities corresponding to the same value of the scale parameter. For large squeezing,
this relation should in fact be determined by Eq.(\ref{eqb1}). The result is shown in Fig.~\ref{fig9}, where the factors
multiplying $u_{rms}/r_d$ and $b_{ho}/r_d$ ($\sqrt{d_i/2}$ and $\sqrt{d_i/(d_i-d_f)}$) have been chosen in such a way that all the curves
follow very much a rather universal curve. Only some of the curves show some discrepancy in the region of very small 
squeezing. In particular this is what happens with potentials II and III in the 3D~$\rightarrow$~1D case, which produce
the bump that differs from the rest of the curves for $\sqrt{d_i/(d_i-d_f)} \langle r_\perp^2\rangle^{1/2}/r_d \approx 1$. This is the same deviation
from the universal curve observed in Fig.~\ref{fig7}c for these two potentials. This universal behaviour is quite well 
reproduced using Eq.(\ref{eqb1}), which, as shown by the dotted line, follows the computed curves almost up to 
the region where the discrepancy mentioned above shows up.

\section{Universal relations}
\label{sec7}

Some of the results shown in the previous section show what we could
consider a universal behaviour. The curves shown in Fig.~\ref{fig7}
are very much independent of the scattering length of the potential,
and of the shape of the potential. Furthermore, the curves shown in
Fig.~\ref{fig8}, and specially the ones in Fig.~\ref{fig9}, can also
be considered independent of the squeezing process.

One could then think that from these universal curves it should be possible to determine the dimension $d$ that
should be used to mimic the squeezing process produced by an external field with squeezing parameter,
$b_{ho}$. This is however not so simple, since, for instance in Fig.~\ref{fig7}, we relate $s$ not just with
$b_{ho}$, but with $b_{ho}/r_d$, and $r_d$ is the root-mean-square radius in the $d$-calculation, Eq.(\ref{rd}), where $d$ must be
the dimension associated to the squeezing parameter, $b_{ho}$. In other words, use of the universal
curves in Fig.~\ref{fig7} to obtain the $s$ value corresponding to some squeezing parameter, $b_{ho}$,
requires previous knowledge of the relation between $d$ and $b_{ho}$. The same
happens in Fig.~\ref{fig8}, where $u_{rms}$ is the root-mean-square radius in the squeezing direction, which can be computed
only after knowing the scale parameter, obtained after maximization of Eqs.(\ref{ov1}) or (\ref{ov2}), which
again requires previous knowledge of the value of $b_{ho}$ associated to a given dimension. The same problem
appears in Fig.~\ref{fig9}.

\begin{figure}[t]
\centering
\includegraphics[width=8.9cm]{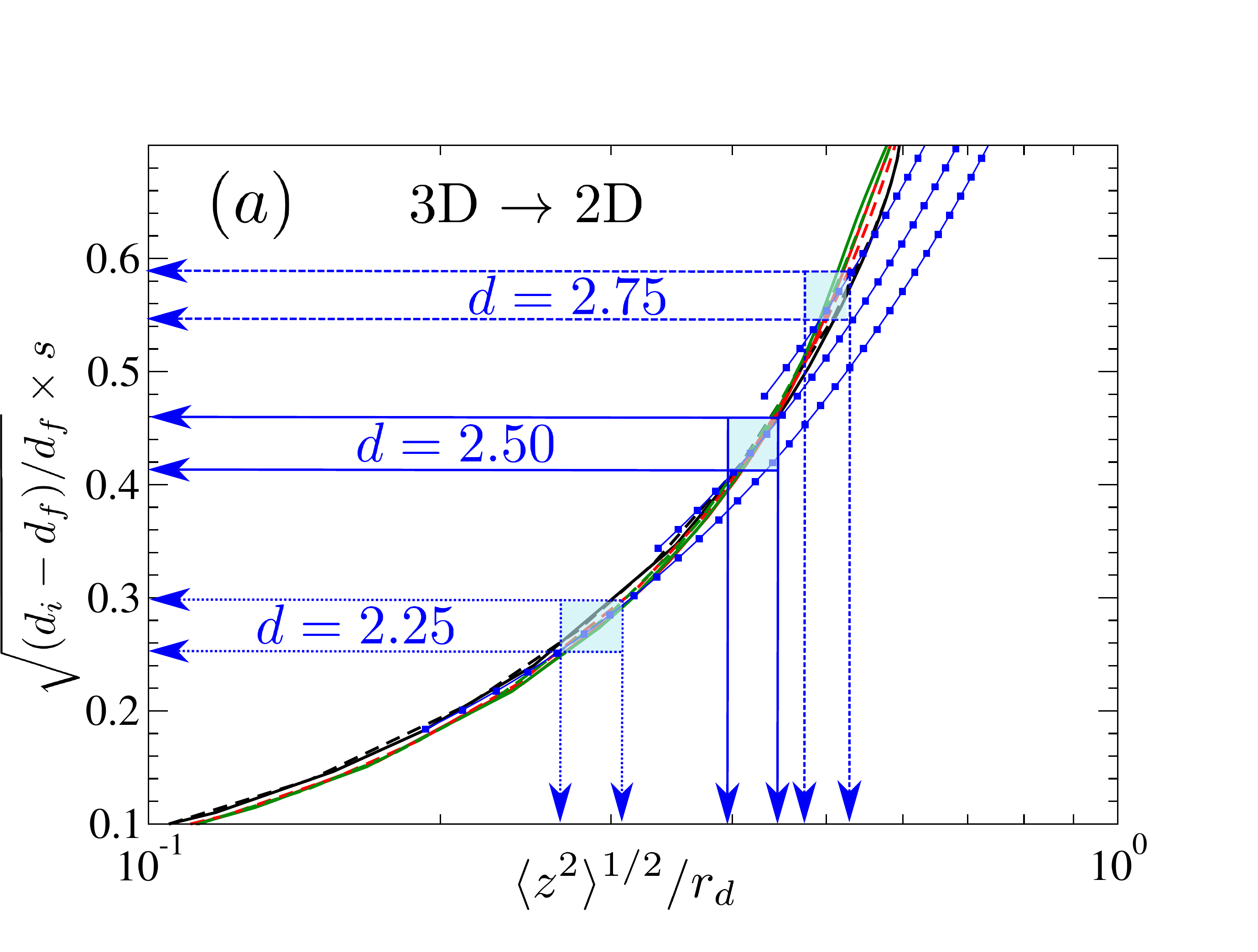}
\includegraphics[width=7.3cm]{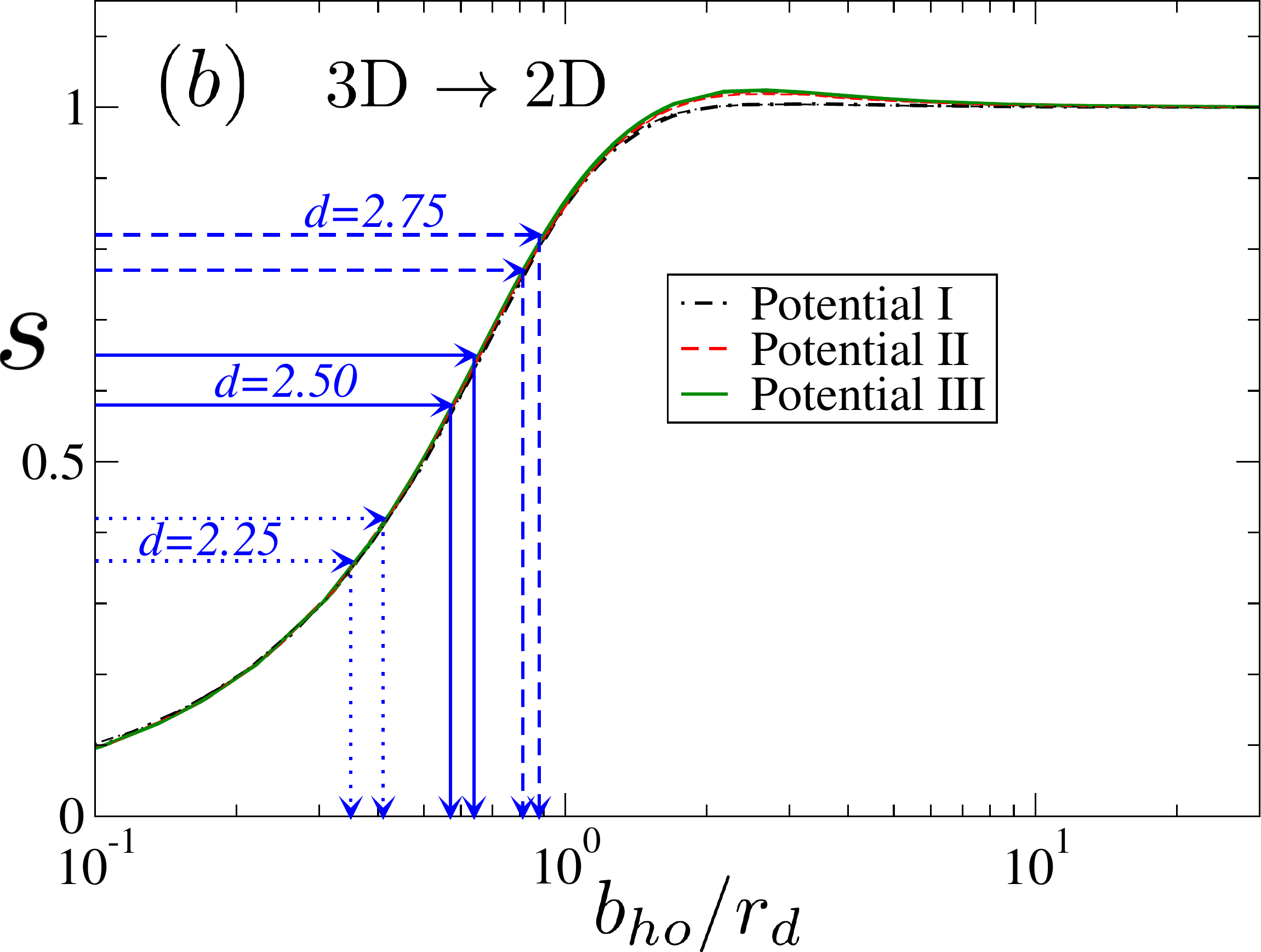}
\caption{$(a)$: The same as in Fig.~\ref{fig8} for the 3D~$\rightarrow$~2D case, where we show (squares) 
the computed values of $\langle z^2 \rangle^{1/2}/r_d$ as a function of the scale parameter $s$ for three different dimensions, $d=2.25$, $2.50$,
and $2.75$. $(b)$: The same as in Fig.~\ref{fig7}b where the arrows indicate the values of $b_{ho}/r_d$ corresponding
to the scale parameter $s$ where the squared lines in the upper part cut the universal curve.}
\label{fig10}
\end{figure}

\begin{table}
\begin{tabular}{c|c|c|c|c}
  $d$  &  $r_d$   &  $\langle z^2\rangle^{1/2}$  &$s$                      & $b_{ho}^b$ \\ \hline
  2.25 &    3.33   &  0.89--1.05 (0.96)                    & 0.36--0.42 (0.39)  & 1.17--1.37  (1.30)\\
  2.50 &    4.24   &  1.69--1.90 (1.77)                    & 0.58--0.65 (0.61) &  2.43--2.69  (2.57)\\
  2.75  &   6.00   &  2.88--3.15  (3.11)                    & 0.77--0.82 (0.81) & 4.92--5.40  (5.38)
\end{tabular}
\caption{Estimate of $\langle z^2\rangle^{1/2}$, the scale parameter $s$, and the squeezing parameter $b_{ho}^b$ for the Morse potential II, obtained from the universal 
curves as shown in Fig.~\ref{fig10} for the dimensions $d=2.25$, $d=2.50$, and $d=2.75$. Their corresponding $r_d$ values are given in the second column of the table.
All the lengths are given in units of the range of the interaction.The numbers within parenthesis are the values obtained from the calculations.}
\label{tab2}
\end{table}

\begin{figure}
\includegraphics[width=7cm]{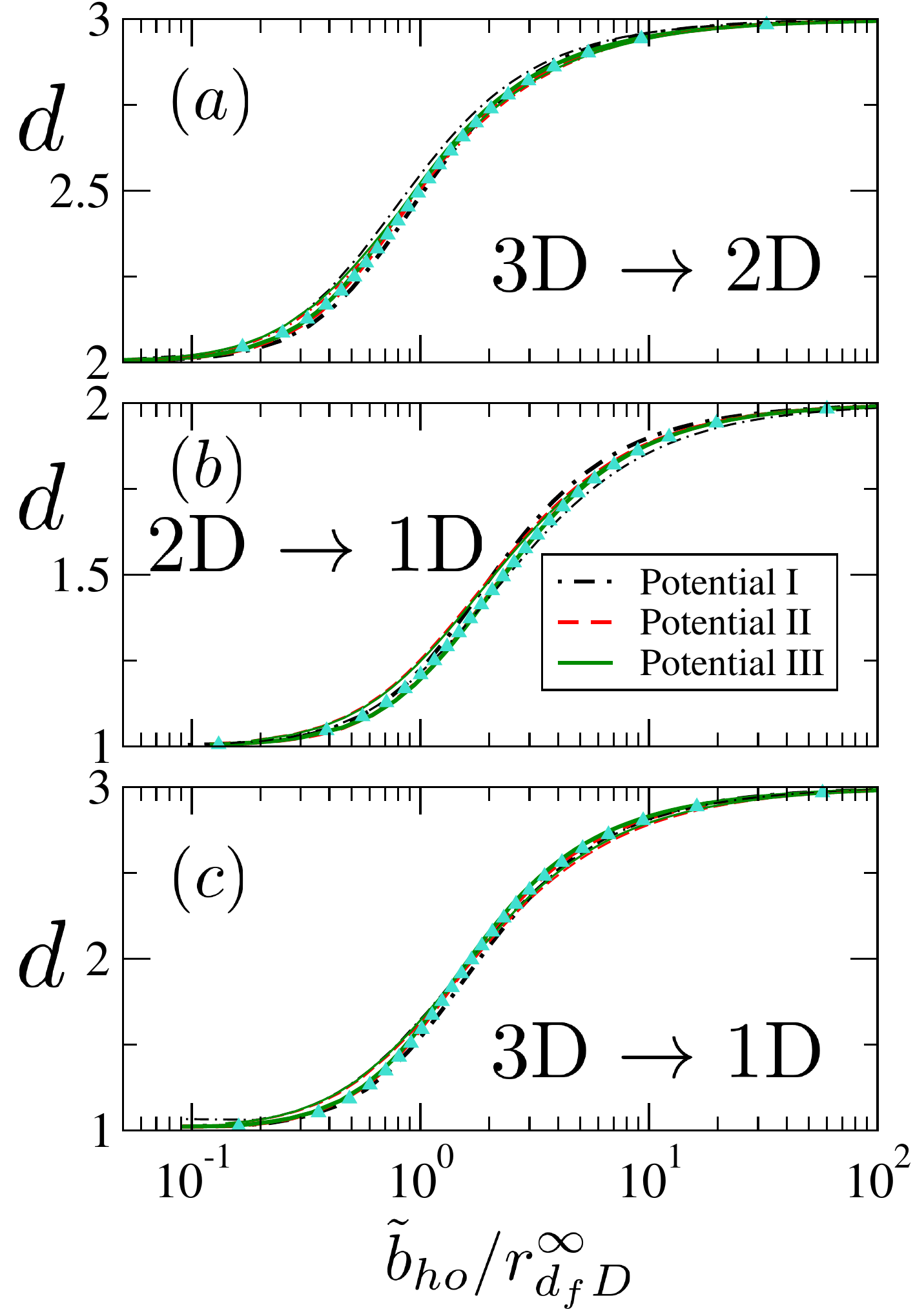}
\caption{The same as Fig.~\ref{fig5}, but after the transformation defined in Eq.(\ref{corr}).}
\label{fig11}     
\end{figure}

However, Figs.~\ref{fig7} to \ref{fig9} can be used to estimate the relation between $d$ and $b_{ho}$ in 
an indirect way. For instance, in Fig.~\ref{fig10}a we show the universal curves shown in Fig.~\ref{fig8}
for the 3D~$\rightarrow$~2D case. On top we plot for three different dimensions, $d=2.25$, $d=2.50$ and $d=2.75$,
the curves (squares) showing $\langle z^2  \rangle^{1/2}/r_d$ as a function of the scale parameter, $s$, for one of the potentials
used in this work, in particular for the Morse potential II. The points where these 
curves cut the universal curve determine the specific values of $s$ and $\langle z^2  \rangle^{1/2}$ corresponding
to each dimension (note that $r_d$ is simply given by Eq.(\ref{rd}), and it does not depend on $s$). Due to the numerical
uncertainty in the universal curve, we have also considered the uncertainty (light-blue rectangles) in where the crossing is actually 
taking place. The results of the estimate for these three dimensions is given in Table~\ref{tab2}, where the second column shows the
$r_d$ value for each dimension, and the third and fourth columns give the estimated range obtained from 
Fig.~\ref{fig10}a for  $\langle z^2  \rangle^{1/2}$ and $s$, respectively. Within parenthesis we give the precise value obtained
from the calculation. As one can see, the estimate is reasonably good.

Once the $s$ value is known, we can use Fig.~\ref{fig7} (or Fig.~\ref{fig9}), as shown in Fig.~\ref{fig10}b, to determine
the values of $b_{ho}$ that correspond to each of the dimensions considered. The results obtained are given in the last column of 
Table~\ref{tab2}, together with the computed values which are given within parenthesis.

In any case, it is obvious that a direct connection between $d$ and $b_{ho}$ is highly desirable. In fact, as shown in Fig.~\ref{fig5}, at least the curves
corresponding to potentials II and III (the ones having a large scattering length) show very much the same behaviour for a given potential shape, but
even if we consider the results with the Gaussian and Morse potentials, the curves are not far of being universal.

An attempt of making the curves in Fig.~\ref{fig5} fully universal was introduced in \cite{gar19}, which can be generalized to a general 
$d_i$D~$\rightarrow$~$d_f$D squeezing process as
\begin{equation}
  \tilde{b}_{ho}=b_{ho} \left(1+\sqrt{\frac{b_{ho}^2+r_{d_fD}^2}{a_{d_i D}^2+r_{d_fD}^2}} \right).
\label{corr}
\end{equation}

The result of this transformation is shown in Fig.~\ref{fig11}. As we can see, the effect of the scattering length being comparable to the range of the potential is 
corrected to a large extent,
and all the curves follow a rather universal curve for each of the $d_i$D~$\rightarrow$~$d_f$D squeezing processes.

\begin{table}
\begin{tabular}{c|c|c|c}
           &  $c_1$   &  $c_2$  & $c_3$                      \\ \hline
  3D~$\rightarrow$~2D  & $-0.28$  &   0.78  &  0.62  \\
  2D~$\rightarrow$~1D  &   1.34     &    0.24  &  0.25 \\
  3D~$\rightarrow$~1D  & $-0.41$ &    1.12  &  0.54     
\end{tabular}
\caption{Parameters used in the numerical fit given in Eq.(\ref{bfit}) giving rise to the curves indicated by the triangles in Fig.~\ref{fig11}. }
\label{tab3}
\end{table}

In Fig.~\ref{fig11} we also show an analytical fit (triangles) that reproduces very well the universal curve
for each of the squeezing scenarios.  Since the curve is model
independent, the special form of the fitting function is unimportant
provided it gives a sufficiently accurate connection
between $\tilde{b}_{ho}$ and $d$. We have different options but one
possibility is
\begin{equation}  
 \frac{\tilde{b}_{ho}}{r_{d_fD}^\infty} = c_1 \left(\frac{d-d_f}{d_i-d}\right)^{d_i/3} + c_2 \tan\left(\left( \frac{d-d_f}{d_i-d_f}\right)^{c_3}  \frac{\pi}{2} \right),
\label{bfit}
\end{equation}
which is a combination of two functions, each of them being equal to zero at $d=d_f$, and to $\infty$
at $d=d_i$, and whose relative weight is used to fit the curves between these two limits.  
The computed fitting constants for each of the three squeezing processes are given in Table~\ref{tab3}.

It is important to mention that, in principle, instead of the two-body energy, as shown in Figs.~\ref{fig3} and \ref{fig4}, one 
could have used a different observable in order to determine the connection between $d$ and the squeezing parameter, $b_{ho}$.
We have checked that when the root-mean-square radii are used, the same universal relation as the one shown in Fig.~\ref{fig11}
is obtained. The main practical problem in this case is that, in general, the convergence of the computed radii when the external squeezing
is considered, is clearly slower than the convergence of the energy. Even larger values of the two-body relative angular momenta
are needed in the expansion (\ref{eq3}), which actually is a source of numerical inaccuracies, especially for large squeezing
scenarios.

\section{Summary and conclusions}
\label{sec8}

We investigate in details how a dimension-dependent centrifugal
barrier can be the substitute for an external one-body potential.  We
choose the ground state of a simple two-body system with Gaussian and
Morse short-range interactions.  The dimension parameter is integer in
the initial formulations, which in this report are analytically
continued to allow non-integer dimensional values.  The external
potential is chosen as the both, experimentally and theoretically,
practical harmonic oscillator which in the present context necessarily
must be anisotropic or deformed.  A well defined unique transformation
between the parameters of the two methods then makes each of them
complete with precise predictions of results from the other method.
The simpler centrifugal barrier computations are then sufficient to
provide observables found with an external potential.

The overall idea is then to start with an ordinary integer dimension
of $3$, $2$, or perhaps $1$, and apply an increasingly confining
external potential in one or more coordinates. This is equivalent to
increasing frequency, or decreasing oscillator length in the
corresponding directions while other coordinates are left untouched.
The process leads from one integer dimension to another lower one.
The results are compared with calculations without external potential
but with a dimension-dependent centrifugal barrier where the same
initial and final configurations are assumed and mathematically
correct.  The aim is to establish a desired unique relation between
the dimension parameter and the oscillator squeezing length.

We first describe in details how the harmonic oscillator confinement
is implemented in the investigated transitions, 3D~$\rightarrow$~2D,
2D~$\rightarrow$~1D and 3D~$\rightarrow$~1D.  The center of mass and
relative coordinates are separated, and the conserved quantum numbers
for the ground states are specified. Second, we discuss properties of
the calculations for the non-integer dimension formulation. A crucial
part is here how to interpret the resulting wave function in terms of
the deformed solution with an external potential. We express how a
scaling of the squeezed coordinate(s) on the ``spherically symmetric''
non-integer dimensional wave function resembles the solution with a
deformed external potential. 

The final results are the unique transformation between the two
methods.  This is for each dimensional transition explicitly given as
a one-to-one correspondence between oscillator squeezing length and
wave-function scaling and dimension parameters.  To be useful it must
be independent of the choice of the short-range interaction. This can be
achieved if the transformation relation is formulated in terms of
either universal quantities or, sufficient for our purpose, quantities
obtainable entirely within the simple dimension calculation.  Both
these cases qualify to be denoted universal relations provided the
results only depend on these quantities, being independent of
dimensional transition and short-range interaction.

In summary, we have first established universal relations.  Second, we
provided the universal interpretation in terms of analytic fitted
functions, which relate oscillator squeezing length with dimension and
wave-function scaling parameters.  These fitted functions allow
predictions from non-integer dimension calculations of observables  in
trap experiments with external potentials.  For two-body systems the
one-to-one correspondence does not provide enormous savings. However,
the idea and the insight obtained through these universal relations
present a new and hopefully useful concept.

In perspectives, the present elaborate report on two-body physics of
non-integer dimensions constitutes the first step in a larger
program.  The immediate next investigations are two-body systems
without bound states in three dimensions.  Then three particles, first
with identical bosons and second for non-identical particles.  These
extensions are each rather big steps presenting their own
difficulties.  In conclusion, we have worked on a simple system to
exhibit new principles, but the door is now open to more complicated
and more interesting systems.

\appendix
\section{The scale parameter and expectation values in the squeezing direction}
\label{app}

\subsection{3D~$\rightarrow$~2D}

In this case the radial coordinate, $r$, is redefined as given in Eq.(\ref{scl1}):
\begin{equation}
r\rightarrow \tilde{r} \equiv \sqrt{x^2+y^2+(z/s)^2} \equiv \sqrt{r_\perp^2+(z/s)^2}.
\end{equation}

The normalization of the wave function (\ref{wf2}) requires calculation of
\begin{equation}
{\cal N}_s=\int r_\perp dr_\perp dz d\varphi |\Psi_d(\tilde{r})|^2,
\label{eqa2}
\end{equation}
which, after defining $u=z/s$, can be rewritten as:
\begin{equation}
{\cal N}_s=s \int r_\perp dr_\perp du d\varphi |\Psi_d(\tilde{r})|^2=s{\cal I}_0,
\label{eqa3}
\end{equation}
where $\tilde{r}^2=r_\perp^2+u^2$ and the integral, ${\cal I}_0$, is independent of the scale
parameter. Note that for $d=3$, since $\Psi_{d=3}$ is already in the 3D space we then
trivially have that ${\cal N}_{s=1}=1$.

Therefore, the wave function, $\tilde{\Psi}_d=\Psi_d/\sqrt{\cal N}_s$, is normalized to 1 in the ordinary
three-dimensional space. After this normalization we can now compute the expectation value, $\langle z^2 \rangle_s$,
which given by
\begin{equation}
\langle z^2 \rangle_s=\int z^2 r_\perp dr_\perp dz d\varphi |\tilde{\Psi}_d(\tilde{r})|^2,
\end{equation}
which, again under the transformation, $u=z/s$, takes the form:
\begin{equation}
\langle z^2 \rangle_s=\frac{s^3}{{\cal N}_s} \int u^2 r_\perp dr_\perp du d\varphi |\Psi_d(\tilde{r})|^2
=\frac{s^3}{{\cal N}_s} {\cal I}_2,
\end{equation}
where ${\cal I}_2$ is independent of $s$.

Making now use of Eq.(\ref{eqa3}) we get:
\begin{equation}
\langle z^2 \rangle_s =s^2 \frac{{\cal I}_2}{{\cal I}_0},
\end{equation}
from which we get the final expression for the scale parameter:
 \begin{equation}
 s=\left( \frac{\langle z^2 \rangle_s }{\langle z^2 \rangle_{s=1} } \right)^{1/2}.
 \label{rat1}
 \end{equation}

 \subsection{3D~$\rightarrow$~1D}

In this case the radial coordinate, $r$, is redefined as given in Eq.(\ref{scl2}):
\begin{equation}
r\rightarrow \tilde{r} \equiv \sqrt{(x^2+y^2)/s^2+z^2} \equiv \sqrt{(r_\perp/s)^2+z^2}.
\end{equation}

We then proceed exactly as in the 3D~$\rightarrow$~2D case, but using the transformation,
$u=r_\perp/s$. In this way the normalization constant (\ref{eqa2}) reads now:
\begin{equation}
{\cal N}_s=s^2 \int u du dz d\varphi |\Psi_d(\tilde{r})|^2=s^2{\cal I}_0.
\label{eqa9}
\end{equation}

In the same way, under the same transformation, the expectation value
\begin{equation}
\langle r_\perp^2 \rangle_s=\int  r_\perp^3 dr_\perp dz d\varphi |\tilde{\Psi}_d(\tilde{r})|^2,
\end{equation}
can be rewritten as:
\begin{equation}
\langle r_\perp^2 \rangle_s=\frac{s^4}{{\cal N}_s} \int  u^3 du dz d\varphi |\Psi_d(\tilde{r})|^2=\frac{s^4}{{\cal N}_s} {\cal I}_2,
\end{equation}
which again, by use of Eq.(\ref{eqa9}) leads to:
 \begin{equation}
 s=\left( \frac{\langle r_\perp^2 \rangle_s }{\langle r_\perp^2 \rangle_{s=1} } \right)^{1/2}.
 \label{rat2}
 \end{equation}

 \subsection{2D~$\rightarrow$~1D}

In this case the radial coordinate, $r$, is redefined as given in Eq.(\ref{scl2}):
\begin{equation}
r\rightarrow \tilde{r} \equiv \sqrt{x^2+(y/s)^2},
\end{equation}
and the normalization constant is given by:
\begin{equation}
{\cal N}_s= \int  dx dy  |\Psi_d(\tilde{r})|^2= s \int  dx du  |\Psi_d(\tilde{r})|^2=s{\cal I}_0,
\end{equation}
where now $u=y/s$.

The expectation value, $\langle y^2 \rangle_s$, is now:
\begin{eqnarray}
\langle y^2 \rangle_s&=& \int  y^2 dx dy  |\tilde{\Psi}_d(\tilde{r})|^2=  \\
& = &\frac{s^3}{{\cal N}_s} \int  u^2 dx du  |\Psi_d(\tilde{r})|^2= s^2 \frac{{\cal I}_2}{{\cal I}_0}. \nonumber
 \end{eqnarray}

 As before, since ${\cal I}_0$ and ${\cal I}_2$ are $s$-independent, we then get the analogous result:
  \begin{equation}
 s=\left( \frac{\langle y^2 \rangle_s }{\langle y^2 \rangle_{s=1} } \right)^{1/2}.
\label{rat3}
 \end{equation}

\acknowledgments 
This work was supported by funds provided by the Ministry of Science, Innovation and University (Spain) under 
contract No. PGC2018-093636-B-I00.

\end{document}